\documentclass[11pt]{article}
 
\usepackage{amsmath,amsthm}
\usepackage{amsthm}
\usepackage{amssymb}
\usepackage{amsfonts} 
\usepackage{color}

\usepackage[latin1]{inputenc}

\usepackage{graphicx}

\newcommand{\sect}[1]{\setcounter{equation}{0}\section{#1}}
\newcommand{\subsect}[1]{\subsection{#1}}

\def\k{{\kappa}}
\newcommand{\sk}{{\rm\ \!S}_\k}            
\newcommand{\ck}{{\rm\ \!C}_\k}           
\newcommand{\tk}{{\rm\ \!T}_\k}

\def\1{\'{\i}}
\newcommand{\uu}{u}

\begin{document}

\bigskip
 
\begin{center}
{\Large{\bf{
The Perlick system type I: from the algebra of symmetries to 
\\[2.5ex]
the geometry of the trajectories}}}
\bigskip 
\bigskip

\begin{center}
\c{S}. Kuru$^a$,  J. Negro$^b$ and O. Ragnisco$^c$
\end{center}

\noindent
$^a$Department of Physics, Faculty of Science, Ankara
University, 06100 Ankara, Turkey

 \noindent
$^b$Departamento de F\'{\i}sica Te\'orica, At\'omica y
\'Optica, Universidad de Valladolid,\\  47011 Valladolid, Spain

\noindent
$^c$Instituto Nazionale di Fisica Nucleare, Sezione di Roma 3,
\\Via della Vasca Navale 84, 00146 Roma, Italy

\end{center}

\bigskip

\begin{abstract}
\noindent 
In this paper, we investigate the main algebraic  properties of the maximally superintegrable system known as ``Perlick system type I". All possible values of the relevant parameters, $K$ and $\beta$, are  considered. In particular, depending on the sign of the parameter $K$ entering in the metrics, the motion will take place on  compact or non compact Riemannian manifolds. To perform our analysis we  follow a classical variant of the so called factorization method. Accordingly, we derive the full set of constants of motion and construct their Poisson algebra. As it is expected for maximally superintegrable systems, the algebraic structure will actually shed light also on the geometric features of the trajectories, that will be depicted for different values of the initial data and of the parameters. Especially, the crucial role played by the rational parameter $\beta$ will be seen ``in action".
\end{abstract}

\bigskip\bigskip 

  


\sect{Introduction}

The Perlick systems type I and II have been studied in a large number of papers, where several crucial features have been discovered (see for instance \cite{Perlick92,PhysD,orlando08,orlando09, AnnPhys2009,orlando11,BEHRRPla,BEHRRAnnPhys,orlando16,IwKatJMP}).
In particular, let us remind that in his original paper \cite{Perlick92}, V. Perlick constructed the most general class of three-dimensional spherically symmetric lagrangian systems complying with the requirements of the celebrated Bertrand's Theorem \cite{bertrand73}, namely  the fact that they possess stable circular orbits and all bounded trajectories are closed. Of course he had to abandon the Euclidean space framework in favour of more general conformally flat manifolds. In this way he unveiled the deep connection existing between the metrics characterizing the manifold and the corresponding integrable  potentials. He classified these generalized Bertrand systems in two multiparametric families, defined according to their Euclidean  limit. Family I is the one yielding the Kepler-Coulomb system and Family II is the one leading to the radial harmonic  oscillator
\cite{ranada05,miller00,miller12}. In ref. \cite{orlando08} by means of the coalgebra approach the extension to hyperspherically symmetric systems in any dimension was performed, and an intrinsic characterization of the two families was proposed. Family I was  denoted as ``intrinsic Kepler-Coulomb" because,  up to additive and multiplicative constants (one could say, ``up to affine transformations"), the pertaining class of potentials were discovered to be just the Green's functions of the corresponding Laplace-Beltrami operators. Analogously, the potentials comprised in Family II, again up to affine transformations, were identified as the ``inverse-squared" of those belonging to Family I and these have been called ``intrinsic oscillator". In a subsequent paper \cite{orlando09} the same authors gave a rigorous proof of the previously obtained results, including that of the periodicity of the bounded motions, and provided a closed expression for the corresponding trajectories. Thus, Perlick's  systems of type I and II are the most general classes of ``maximally superintegrable" lagrangian (and consequently Hamiltonian) systems with (hyper-)spherical symmetry.

\noindent

To further clarify our previous statements, let us recall that a classical Hamiltonian system with $n$
degrees of freedom is said to be integrable if  there are $n$ functionally independent globally defined and single-valued integrals of motion in involution for a given Poisson bracket. If there exist
$k$,  with $0<k\leq n-1$, additional integrals (they must be functionally independent, but not all of them in involution) it is called superintegrable. When $k=n-1$, the system is maximally superintegrable. In this case, all bounded trajectories are closed and the motion is periodic
\cite{woj83,evans90,winternitz13}. The constants of motion of maximally superintegrable systems 
close an algebra and can be used to determine the trajectories in pure algebraic way.  Kepler-Coulomb and (isotropic) harmonic oscillator systems 
are examples of spherically symmetric and maximally superintegrable systems
 in an $n$ dimensional Euclidean space. 
In dealing here with a maximally superintegrable system such as Perlick system of type I, rather than relying on an analytic  approach one as in \cite{orlando09},  we will follow  an algebraic one,  that can be generally called ``factorization approach" or ``factorization method". 
Through the factorization method the algebra of the integrals  of motions can  be explicitly constructed, and its features will pave the way to identifying relevant geometric properties such as  the shape of the closed orbits.
We wish to remark that the factorization method applies both  to classical and
quantum  systems. The method is quite simple 
in  yielding the  constants of motion (symmetries) and  unveiling
their underlying algebra \cite{negro08,hussin11,negro12,negro13,negro14,negro16,negro17,TaubSGA}. We should mention that other methods have been successfully applied 
to find the constants of motion of superintegrable systems. One of them is based on the Hamilton-Jacobi equation, 
as it is shown in many references \cite{miller12,winternitz10,hakobyan12}. Another interesting way to look for constants of motion
is the method of coalgebras developed in \cite{riglioni13,riglioni15}. A systematic
algebraic method has also been used in other papers \cite{ian07,ian13}. In this work, we will  perform a systematic investigation of the constants of motion of the  classical superintegrable three dimensional Perlick system type I  by using the factorization method in order to derive the trajectories in a purely  algebraic way.

This paper is organized as follows.
In section 2,  we briefly recall the basic definition of the Perlick system of type I which
depends essentially on two parameters $\beta$ and $K$. We will introduce new variables more
suitable to describe the compact or non compact manifolds where this system may be defined.
In section 3, we construct the constants of motion  by means of an extension of the factorization
method to classical three dimensional systems (it can be applied to higher dimensions too). 
The trajectories of Perlick I system are derived, discussed and depicted, for different values of the initial conditions as well as of the relevant parameters of the system, i.e, $K$ 
 and $\beta = m/n$. Both open and closed trajectories are  presented.   
The crucial point of the whole treatment, namely the evaluation of the algebra of the constants of motion is also given here.
Section 4 is devoted to the special case of the motion in the plane. The key role played by the requirement that $\beta$ be a rational number becomes  extremely clear in this somehow simplified setting. 
Finally,  in  section 5, we summarize the new results that we have obtained, emphasizing what are in our opinion the most relevant ones. As a closing comment, we will briefly outline the most promising developments.

\sect{The Perlick system type I}
The Hamiltonian for the Perlick system type I in spherical 
coordinates $(r,\theta,\varphi)$ is
\begin{equation}\label{hperlik1}
\widetilde{H}^\pm=\beta^2 (1+K\, r^2)\,\dfrac{p_r^2}{2}+\dfrac{{\bf L}^2}{2\,r^2}+G
\pm\dfrac{1}{r}\,\sqrt{1+K\,r^2}\,,
\end{equation}
where ${\bf L}$ is the angular momentum and ${\bf L}^2$ can be considered as an angular Hamiltonian $H_{\theta \varphi}$ in the variables $(\theta,\varphi)$, having the form
\begin{equation}\label{hperlikL}
{\bf L}^2=H_{\theta \varphi}=p_{\theta}^2+\dfrac{p_\varphi^2 }{\sin^2{\theta} }\,
\end{equation}
being $K$, $G$ real constants and $\beta=m/n$ a rational number.  
We can also consider $p_\varphi^2$ as a free Hamiltonian $H_\varphi$ 
in the variables $(\varphi,p_\varphi)$, defined on the unit circle. The angular variables
have the usual range: $0\leq \theta \leq \pi$ and $0 \leq \varphi<2\pi$. The radial variable
has different ranges depending on $K$. If $K\geq0$, then  $0< r < \infty$, while
if $K<0$, $r$ must be restricted to the finite interval $0< r < 1/\sqrt{-K}$. In his paper
\cite{Perlick92} Perlick considered only the negative sign in front of
the square root of the potential, that is $\widetilde{H}^-$, in order to have bounded
motions. However, we will
show that for the case $K<0$ both Hamiltonians $\widetilde H^\pm$ should be taken 
into account in order to have a well defined Hamiltonian on a three dimensional
sphere. The metric associated to this Hamiltonian is \cite{orlando08}
\begin{equation}\label{metric}
ds^2=\frac{dr^2}{\beta^2\,(1+K\,r^2)}+r^2\,(d\theta^2+\sin^2\theta\,d\varphi ^2)\,.
\end{equation}
Another point of view on the Hamiltonians (\ref{hperlik1}) is to look
at them not as defined in a curved space but as position dependent mass
Hamiltonians in a flat space \cite{quesne04,ballesteros17}.

Since $\{{\bf L}^2,\widetilde H^\pm\}=0$,  ${\bf L}^2$ is a constant of motion and  takes the positive values  ${\bf L}^2=\ell^2$. 
Here, $\{\cdot,\cdot\}$ stands  for the Poisson brackets (PBs) that for the  functions $f(r, \theta, \phi), \,g(r, \theta, \varphi)$ in spherical coordinates are defined by
\begin{equation}\label{pbs}
\{f,g\}=\dfrac{\partial f}{\partial r}\dfrac{\partial g}{\partial p_r}-\dfrac{\partial f}{\partial p_r}\dfrac{\partial g}{\partial r}+\dfrac{\partial f}{\partial \theta}\dfrac{\partial g}{\partial p_\theta}-\dfrac{\partial f}{\partial p_\theta}\dfrac{\partial g}{\partial \theta}+\dfrac{\partial f}{\partial \varphi}\dfrac{\partial g}{\partial p_\varphi}-\dfrac{\partial f}{\partial p_\varphi}\dfrac{\partial g}{\partial \varphi}\,.
\end{equation}
For the sake
of simplicity, we also choose $G=0$, so when we replace ${\bf L}^2$ by its constant value $\ell^2$
we can rewrite the initial Hamiltonian (\ref{hperlik1}) as an effective Hamiltonian $\widetilde H_r^\pm$ in the variable $r$:
\begin{equation}\label{hperlik2}
\widetilde H_r^\pm=\beta^2 (1+K \,r^2)\,\dfrac{p_r^2}{2}+
\dfrac{\ell^2}{2\,r^2}\pm\dfrac{1}{r}\,\sqrt{1+K\,r^2}\,.
\end{equation}
Since $\{p_{\varphi},\widetilde H^\pm\}=0$, $p_{\varphi}$ is another constant of motion: $p_{\varphi}=\ell_z=const$. 
In the same way, after replacing $H_{\varphi}$ by $\ell_z$ in (\ref{hperlikL}), we have an effective Hamiltonian $H_\theta$ in $\theta$:
\begin{equation}\label{hperlikL1}
H_{\theta}=p_{\theta}^2+\dfrac{\ell_z^2 }{\sin^2{\theta} }\,.
\end{equation}
Notice that $H_{\theta}$ is singular at the angles $\theta=0$ and $\theta=\pi$, so the trajectories lie between these two values (i.e. the North and South poles cannot be reached unless $\ell_z=0$).
Once fixed the values of $\ell, \ell_z$, the turning points for $\theta$ are given by the solutions of $\ell^2={\ell_z^2 }/{\sin^2{\theta} }$. 

In order to take into account different signs and values of $K$,
we will make use of the $\k$-dependent cosine and sine functions defined  
by 
\begin{equation}
\ck(\uu)\equiv\left\{
\begin{array}{ll}
  \cos {\sqrt{\k}\, \uu} &\quad  \k>0 \\ 
\qquad 1  &\quad
  \k=0 \\
\cosh {\sqrt{-\k}\, \uu} &\quad   \k<0 
\end{array}\right.  ,
\quad
   \sk(\uu)
\equiv\left\{
\begin{array}{ll}
  \frac{1}{\sqrt{\k}} \sin {\sqrt{\k}\, \uu} &\quad  \k>0 \\ 
\qquad \uu  &\quad
  \k=0 \\ 
\frac{1}{\sqrt{-\k}} \sinh {\sqrt{-\k}\, \uu} &\quad  \k<0 
\end{array}\right.  .
\end{equation}
The $\k$-tangent is defined by 
$$
 \tk(\uu)\equiv\frac{\sk(\uu)}{\ck(\uu)}\,.
 $$
Some  relations among these $\k$-functions are \cite{conf,mariano99}: 
\begin{equation}
\ck^2(\uu)+\k\sk^2(\uu)=1,\qquad  
 \frac{ {\rm d}}{{\rm d} \uu}\ck(\uu)=-\k\sk(\uu),\qquad         \frac{ {\rm d}}
{{\rm d} \uu}\sk(\uu)= \ck(\uu)\,.
\label{za}
\end{equation}

Now, let us introduce the parameter $\kappa$ instead of $K$ and make the following change of canonical variables, 
\begin{equation}\label{canvariable}
K=-\kappa\,,\qquad r=\sk(\xi),\qquad p_r=\frac{p_\xi}{\ck(\xi)}\,,
\end{equation}
where the range of $\xi$ is as follows: (a) for $\k \leq 0$, $0<\xi<\infty$;
(b) for $\kappa>0$, we have two natural choices: $0<\xi<\pi/(2\sqrt{\k})$
or $\pi/(2\sqrt{\k})<\xi<\pi/\sqrt{\k})$. This change of coordinates can be interpreted in the following way. 
The variables $(\xi,\theta,\varphi)$
can be considered as the angular coordinates of the points of
a (pseudo) sphere in $\mathbb R^4$. When $\k<0$ they parametrize the points of one sheet of
a three dimensional (3D) hyperboloid; when $\k=0$ the coordinates $(\xi,\theta,\varphi)$ 
represent the spherical coordinates of the points of $\mathbb R^3$; finally, 
for $\k>0$ these variables with
$0<\xi<\pi/(2\sqrt{\k})$ parametrize the points of one half (the north hemisphere) 
of a deformed 3D sphere, and the values $\pi/(2\sqrt{\k})<\xi<\pi/\sqrt{\k}$ give the points of the south hemisphere. In the case $\k<0$, this choice of coordinates 
is not important since the north and south
hemispheres of a hyperboloid are not connected and can be taken independently
without any problem, but in the case $\k>0$ this parametrization of the points of a sphere
is quite relevant.

Next, we introduce a Hamiltonian $H$ in the variables $(\xi,\theta,\varphi)$ defined as follows
for any value of $\k$:
\begin{equation}\label{hperlik33a}
H(\xi,\theta,\varphi)=\beta^2 \,\dfrac{p_\xi^2}{2}
+\dfrac{{\bf L}^2}{2}\frac{1}{\sk^2(\xi)}-\frac{1}{\tk(\xi)}\,,
\end{equation}
and the corresponding effective Hamiltonian $H_\xi$ by  
\begin{equation}\label{hperlik3a}
H_\xi=\beta^2 \,\dfrac{p_\xi^2}{2}
+\dfrac{{\ell}^2}{2}\frac{1}{\sk^2(\xi)} -\frac{1}{\tk(\xi)}=\beta^2 \,\dfrac{p_\xi^2}{2}+V_{\rm eff}({\xi})\,.
\end{equation}

We will see the relation of this Hamiltonian with the Perlick Hamiltonians $\widetilde H^\pm$
for different values of $\k$.
\begin{itemize}

\item $\k \leq 0$. In this case, both variables $\xi$ and $r$ vary in the positive semi-line 
$(0,\infty)$. If we change the variables
in (\ref{hperlik1}) according to (\ref{canvariable})
it is easy to check that 
\begin{equation}
H(\xi,\theta,\varphi)=\widetilde H^-(r,\theta,\varphi) \,.
\end{equation}

\item $\k>0$. Here, there are two complementary intervals for the range of variable $\xi$ as mentioned above.
If we change the variables
in (\ref{hperlik1}) according to (\ref{canvariable})
we get 
\begin{equation}
H(\xi,\theta,\varphi) =
\left\{\begin{array}{ll}
\widetilde H^-(r,\theta,\varphi)  \,,
\qquad & 0<\xi<\dfrac{\pi}{2\sqrt{\k}}\,,\qquad 0<r<\dfrac{1}{\sqrt{\k}},
\\[2.5ex]
\widetilde H^+(r,\theta,\varphi)  \,,
\qquad & \dfrac{\pi}{2\sqrt{\k}}<\xi<\dfrac{\pi}{\sqrt{\k}}\,,\qquad \dfrac{1}{\sqrt{\k}}>r>0\,.
\end{array}
\right.
\end{equation}
Therefore, in this case we have a unique Hamiltonian $H$ well defined for all the points
of the sphere $ 0<\xi<{\pi}/{\sqrt{\k}}$, such that on the north hemisphere is described by $\widetilde H^-$ and on the south by
$\widetilde H^+$. When these points are parametrized in terms of $(r,\theta,\varphi)$
variables, each hemisphere gives rise to the two distinct Hamiltonians 
$\widetilde H^\pm$. This is very important since in general the trajectories on the sphere are not restricted to the upper or the lower hemispheres,
and a full description of all of them should be done by means of $H$.

\end{itemize}

Henceforth, we will use the following notation. For all the values of $\k$ we will 
take the Hamiltonian $H(\xi,\theta,\varphi)$ as given in (\ref{hperlik33a})
or the effective Hamiltonian $H_\xi(\xi)$ in (\ref{hperlik3a}). The ranges of the 
variable $\xi$ are taken according to the previous discussion depending on
the values of $\k$, in particular for $\k>0$ the interval is $0<\xi<\frac{\pi}{\sqrt{\k}}$.

\sect{Constants of motion from factorization properties}

In this model, we already have three independent constants of motion:
$H, H_{\theta \varphi}, H_\varphi$ with constant values given by $E,\ell^2,\ell_z^2$.
Our aim is to find additional constants of motion by means of the factorization method in order
to show the maximal superintegrability of the system.
 A first pair of constants $X^\pm$ will be obtained
from  the effective  Hamiltonians $H_\xi$ and $H_\theta$, and a second pair $Y^\pm$ from the next two effective Hamiltonians, 
$H_\theta$ and $H_\varphi$.

\subsect{The constants of motion $X^\pm$}
The first set of constants $X^\pm$ are constructed in terms of the shift functions 
of $H_\xi$ and the ladder functions of $H_\theta$ which are obtained
as follows \cite{negro08}.

\begin{itemize}
\item
{\it Shift functions of $H_\xi$}

The Hamiltonian (\ref{hperlik3a}) can be factorized as
\begin{equation}\label{hfac}
H_\xi=B^+ B^-+\lambda_\xi\,,
\end{equation}
where
\begin{equation}\label{hfacb}
B^{\pm}= \frac{1}{\sqrt{2}} \,\left(\mp\,i\,{\beta }\,{p_\xi}+\frac{\ell}{\tk{(\xi)}}- \dfrac{1}{\ell}\right)\,,\qquad \lambda_\xi=-\dfrac12\left(\dfrac{1}{\ell^2}-\k\,\ell^2\right)\,.
\end{equation}
The shift functions $B^\pm$ are complex conjugate of each other and they satisfy the following PBs together with the Hamiltonian $H_\xi$
\begin{equation}\label{hbpb}
\{B^-,B^+\}=i\,\beta\,\dfrac{\ell}{{\sk}^2 (\xi)}\,,
\qquad\{H_\xi,B^{\pm}\}=\pm i\,\beta\,\dfrac{\ell}{{\sk}^2 (\xi)}\, B^{\pm}\,.
\end{equation}
The second PB of (\ref{hbpb}) implies that
\begin{equation}\label{hbpb2}
\{H,B^{\pm}\}=\pm i\,\beta\,\dfrac{\sqrt{H_{\theta \varphi}}}{{\sk}^2 (\xi)}\, B^{\pm}\,.
\end{equation}

From the factorization given in (\ref{hfac}) or the effective potential (\ref{hperlik3a}), we conclude that the energy $E$ of the
total Hamiltonian $H_\xi$ for bounded motions must satisfy the following inequalities depending on $\k$:
\begin{equation}\label{energycondition}
\begin{array}{ll}
\k<0\,,
\qquad &-\sqrt{|\k|}>E\geq -\dfrac12\left(\dfrac1{\ell^2}+|\k| \ell^2\right)\,,
\\ [2.ex]
\k=0\,,
\qquad & 0>E\geq -\dfrac1{2\,\ell^2}\,,
\\[2.ex]
\k>0\,,
\qquad &\infty>E\geq -\dfrac12\left(\dfrac1{\ell^2}-\k \ell^2\right)\,.
\end{array}
\end{equation}
In Fig.~\ref{effectivepot}, it is shown the effective potential $V_{\rm eff}({\xi})$ given by (\ref{hperlik3a}) together with the conditions on the energy (\ref{energycondition}) for different values of $\k$. For all these three cases, bounded motions have two turning points in the variable $\xi$ given by  $E=V_{\rm eff}({\xi})$.
For $E\geq -\sqrt{|\k|}$ the trajectory will be unbounded with only one turning point for $\k \leq 0$.

\begin{figure}
\centering
\includegraphics[width=0.31\textwidth]{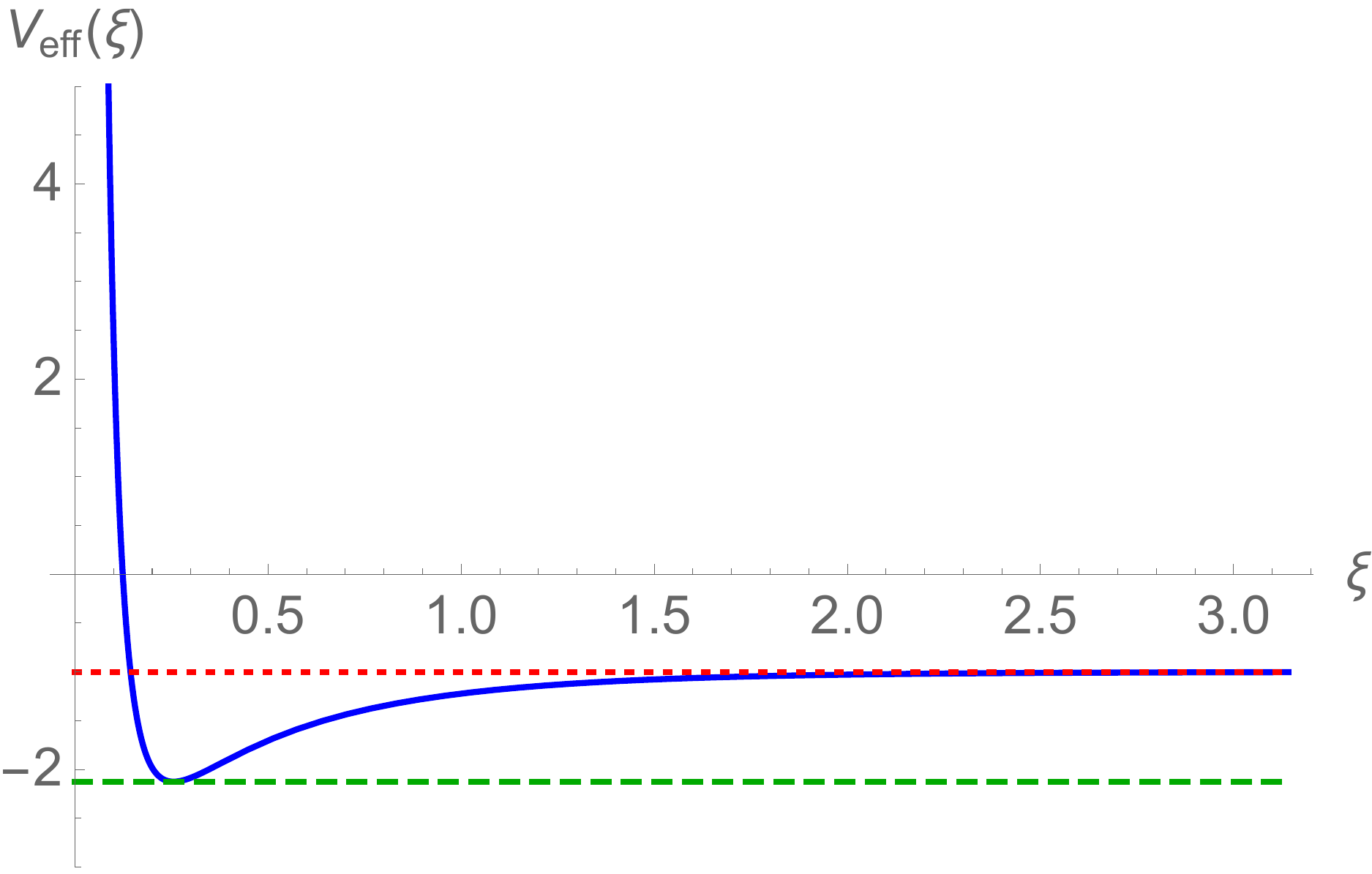}\quad
\includegraphics[width=0.31\textwidth]{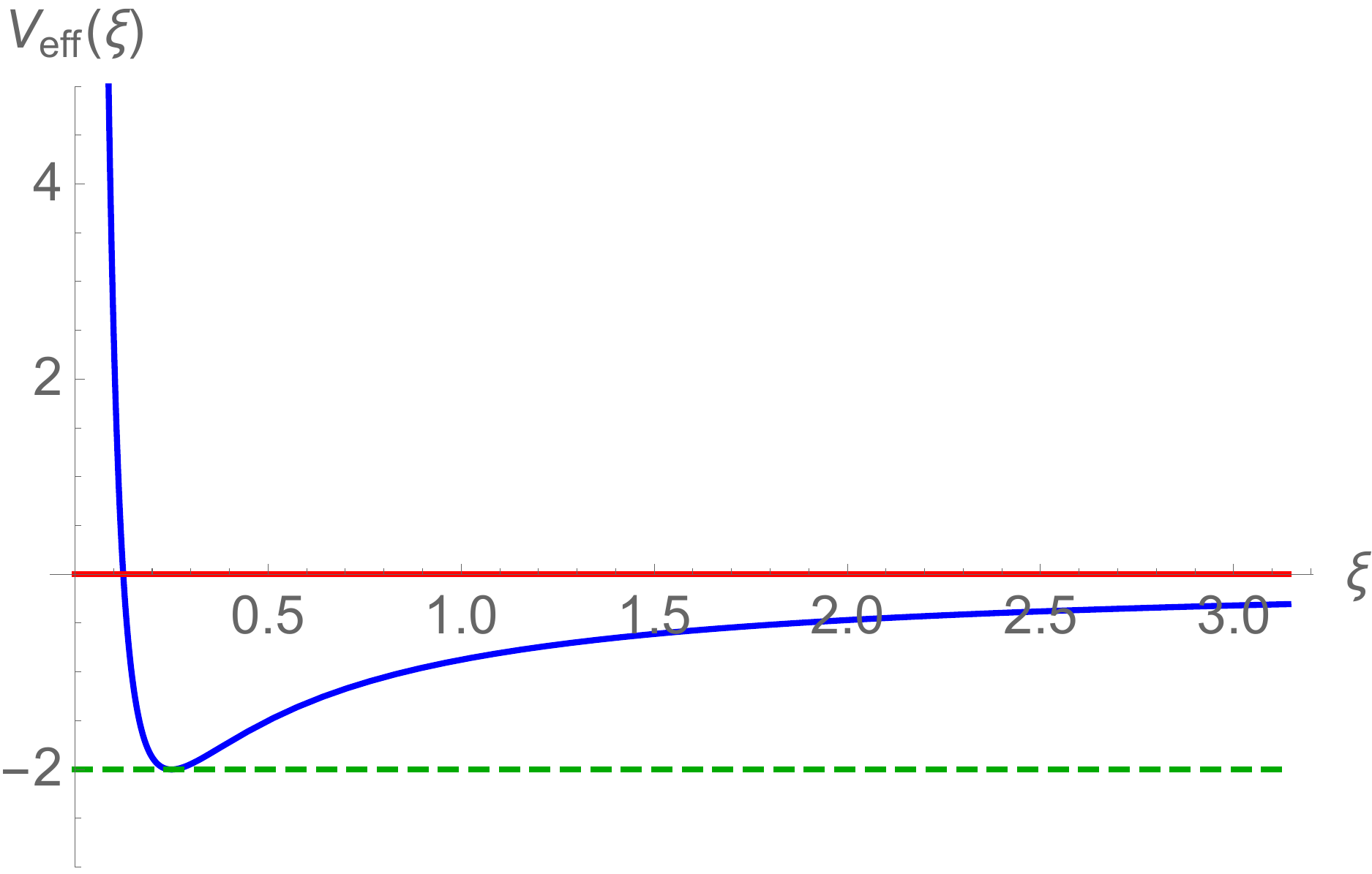}\quad
\includegraphics[width=0.31\textwidth]{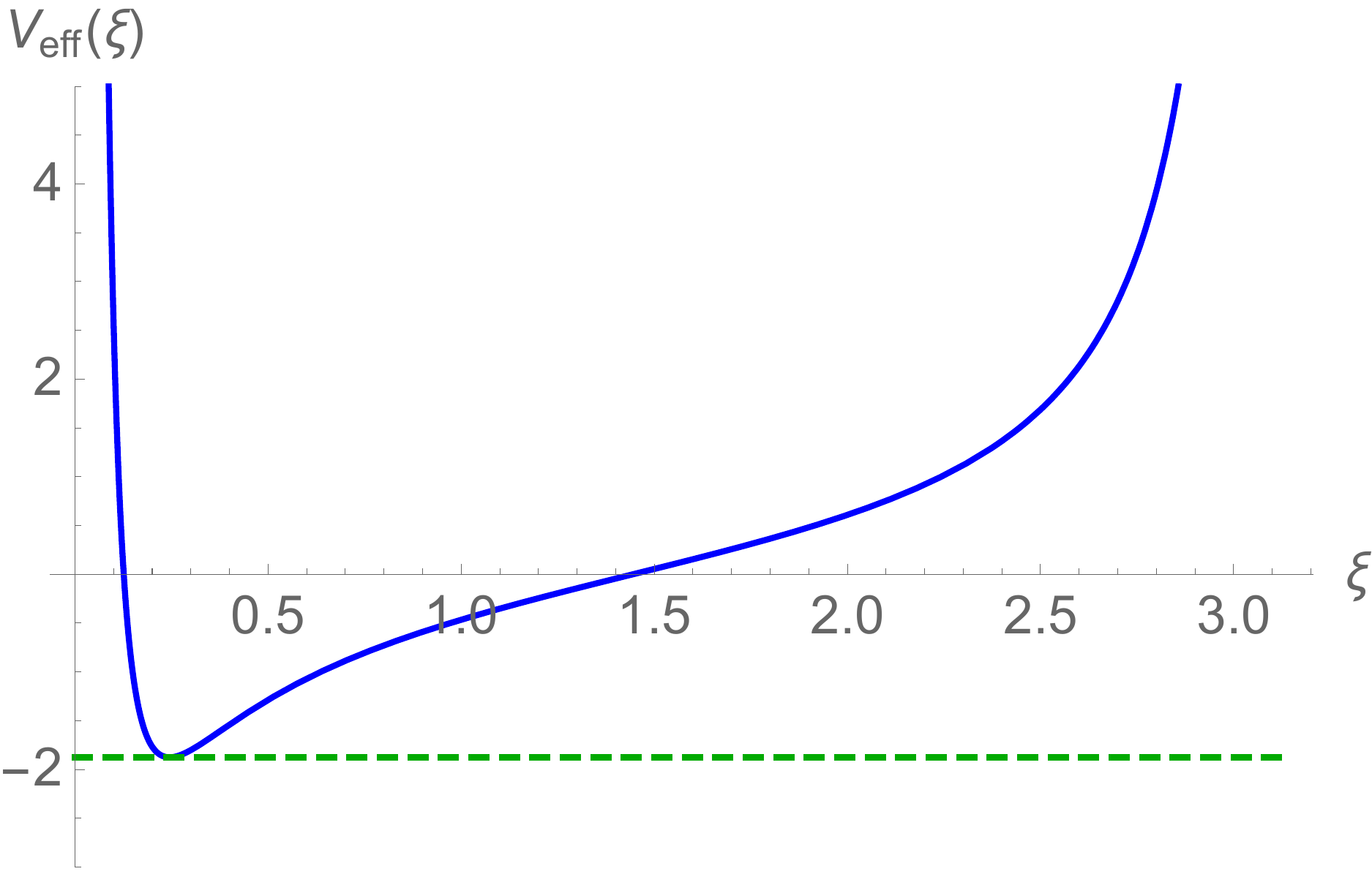}
\caption{\small Plot of the effective potential  $V_{\rm eff}({\xi})$ for $\k=-1$ (left), $\k=0$ (center) and $\k=1$ (right) together with the limiting values of the energy given by the inequalities (\ref{energycondition}) for $\ell=0.5$.  The range of $\k$ for $\k \leq 0$ is $0<\xi<\infty$ and for $\kappa>0$ it is  $0<\xi<\pi/\sqrt{\k}$.
\label{effectivepot}}
\end{figure}

\item
{\it Ladder functions of $H_{\theta}$}

In order to find the ladder functions for the angular Hamiltonian $H_\theta$, defined by (\ref{hperlikL1}), first we multiply it by  ${\sin^2{\theta} }$ and after rearranging, we get 
\begin{equation}\label{hperlikL2}
-{\ell_z^2 }=p_{\theta}^2\,{\sin^2{\theta} }-H_{\theta}\,{\sin^2{\theta} }\,.
\end{equation}
Now, we can factorize the right hand side of this equality in terms of complex conjugate ladder functions
\begin{equation}\label{htfac}
-\ell_z^2=A^+ A^-+\lambda_{\theta}\,,
\end{equation}
where
\begin{equation}\label{htfaca}
A^{\pm}=\mp i \,\sin{\theta}\,{p_{\theta}}+\sqrt{H_{\theta}} \cos{\theta}\,,\qquad \lambda_{\theta}=-H_{\theta}\,.
\end{equation}
These ladder functions $A^{\pm}$ together with the Hamiltonian $H_{\theta}$ satisfy the following PBs
\begin{equation}\label{hbpa}
\{A^-,A^+\}=2\,i\,\sqrt{H_{\theta}}\,,\qquad\{H_{\theta},A^{\pm}\}=\pm2\, i\,\sqrt{H_{\theta}}\, A^{\pm}\,.
\end{equation}
Therefore, the PB with $H$ is
\begin{equation}\label{hbpax}
\{H,A^{\pm}\}=\mp i\,\dfrac{\sqrt{H_{\theta \varphi}}}{{\sk}^2 (\xi)}\, A^{\pm}\,.
\end{equation}
Notice  that the factorization (\ref{htfac}) implies the following inequality:
\begin{equation}
\ell^2\geq \ell^2_z \,,
\end{equation}
which is reasonable in view of the physical interpretation of $\ell$ (the total angular momentum) and $\ell_z$ (the component of the angular momentum in the $z$-direction).

\item
{\it Construction of $X^\pm$}

From the PBs (\ref{hbpb2}) and (\ref{hbpax}), taking into account that $\beta=m/n$, we obtain the constants of motion $X^{\pm}$ of the Perlick system type I in the following form
\begin{equation}\label{X}
\{H,X^{\pm}\}=0\,,\qquad X^{\pm}=(A^{\pm})^m(B^{\mp})^n\,.
\end{equation}
We can denote the values of these constants as follows,
\begin{equation}\label{xpm}
X^\pm = q_{\rm  x}\, e^{\pm i\, \alpha_{\rm  x}},\qquad 0\leq q_{\rm  x}<\infty,\ \  -\pi \leq \alpha_{\rm  x}<\pi.
\end{equation}
The absolute value $q_{\rm  x}$ is directly obtained
from the factorization properties of $A^\pm$ and $B^\pm$:
\begin{equation}
q_{\rm  x}=|X^{\pm}| = 
\left(\ell^2-\ell_z^2\right)^{m/2}
\left(E+\frac12\left(\frac1{\ell^2}-\k\, \ell^2\right)\right)^{n/2} \, .
\end{equation}

\end{itemize}

By means of these constants of motion, the relation between the variables $\xi$ (or $r$) and $\theta$ 
along the trajectories can be found from (\ref{xpm}) (and the relative frequencies as we will see later). However, when $\ell=\ell_z$, this relation breaks,
because in this case (\ref{hperlikL1}) and (\ref{htfaca}) entail $p_\theta=0$ and $\theta= \pi/2$. In other words, if $\ell=\ell_z$, the motion
takes place in the horizontal plane $z=0$. This particular case will be considered in
Section~4. 

In principle, the constants of motion $X^\pm$ given by (\ref{X}) are not polynomial
in the momentum functions,
since $A^\pm$ and $B^\pm$ depend on $\ell$ which is a square root (see  (\ref{hperlikL1})). However,
if we expand the powers $m$ and $n$ in (\ref{X}), then the real and imaginary
parts of these functions will give rise to polynomial constants of motion in the momenta \cite{negro14}.

\subsect{The constants of motion $Y^\pm$}

This new set of constants of motion is derived from the shift functions
of $H_\theta$ and the ladder functions of $H_\varphi$. They
are obtained in a similar way as the previous set.

\begin{itemize}
\item
{\it Shift functions of $H_{\theta}$}

The angular Hamiltonian defined by (\ref{hperlikL1}) is factorized as
\begin{equation}\label{hperlikL2fac}
H_{\theta}=p_{\theta}^2+\dfrac{\ell_z^2 }{\sin^2{\theta} }=C^+ C^-+\lambda_{\ell}\,, 
\end{equation}
where
\begin{equation}\label{htfacc}
C^{\pm}=\mp i \,{p_{\theta}}+\ell_z \cot{\theta}\,,\qquad \lambda_{\ell}=\ell_z^2\,.
\end{equation}
These factor functions $C^{\pm}$ together with the Hamiltonian $H_{\theta}$ satisfy the following PBs
\begin{equation}\label{hbpc}
\{C^-,C^+\}=2\,i\,\frac{\ell_z}{\sin^2{\theta}}\,,\qquad\{H_{\theta},C^{\pm}\}=\pm2 \,i\,\frac{\ell_z}{\sin^2{\theta}}\, C^{\pm}\,.
\end{equation}
The second PB of (\ref{hbpc}) implies that
\begin{equation}\label{hbpc2}
\{H_{\theta \varphi},C^{\pm}\}=\pm2 \,i\,\frac{\sqrt{H_{\varphi}}}{\sin^2{\theta}}\, C^{\pm}\,.
\end{equation}

\item
{\it Ladder functions of $H_{\varphi}$}

The Hamiltonian $H_{\varphi}=p_{\varphi}^2$ is factorized as
\begin{equation}\label{hphiladder}
H_{\varphi}=D^+ D^-\,, \qquad D^{\pm}=\sqrt{H_{\varphi}}\,e^{ \mp\,i \,{\varphi}}\,.
\end{equation}
These ladder functions $D^{\pm}$ together with the Hamiltonian $H_{\varphi}$ satisfy the following PBs
\begin{equation}\label{hbpd}
\{D^-,D^+\}=2\,i\,\sqrt{H_{\varphi}},\qquad
\{H_\varphi, D^\pm\}= \pm 2 \,i\,\sqrt{H_{\varphi}}\, D^\pm\,.
\end{equation}
Then, the PB with the Hamiltonian $H_{\theta \varphi}$ is
\begin{equation}\label{hbpd2}
\{H_{\theta \varphi},D^{\pm}\}=\pm2 \,i\,\frac{\sqrt{H_{\varphi}}}{\sin^2{\theta}}\, D^{\pm}\,.
\end{equation}

\item
{\it The building of $Y^\pm$}

The Hamiltonian $H_{\theta \varphi}$ defined by (\ref{hperlikL1}), besides $H_\varphi$, has the constants of motion $Y^\pm$ that now can be found with the help of the relations (\ref{hbpc2}) and (\ref{hbpd2}):
\begin{equation}\label{Y}
\{H_{\theta \varphi},Y^{\pm}\}=0\,,\qquad Y^{\pm}=(C^{\pm})(D^{\mp})\,.
\end{equation}
The geometric meaning of these constants of motion can be appreciated by rewriting them in the
form
\begin{equation}\label{ypm2}
Y^\pm =- L_z(L_x\pm i L_y)\,,
\end{equation}
where $L_x$, $L_y$ and $L_z$ are the components of the angular momentum
in the $x$, $y$  and $z$ direction, respectively:
\begin{equation}\label{lxly}
L_x=-\sin{\varphi}\, p_{\theta}-\cot{\theta}\,\cos{\varphi}\, p_{\varphi}\,,\quad L_y=\cos{\varphi}\, p_{\theta}-\cot{\theta}\,\sin{\varphi}\, p_{\varphi}\,,\quad L_z=p_{\varphi}\,.
\end{equation}

If we denote the values of these constants of motion by
\begin{equation}\label{ypm}
Y^\pm = q_{\rm y}\, e^{\pm i \alpha_{\rm y}},\qquad 0\leq q_{\rm y}<\infty,
\quad \pi \leq \alpha_{\rm y}<\pi\,,
\end{equation}
with 
\begin{equation}
q_{\rm y}= |Y^\pm| = \ell_z(\ell^2-\ell_z^2)^{1/2}\,,
\end{equation}
while the angle $\alpha_{\rm y}$ coincides with the azimuthal angle of the
angular momentum vector $\boldsymbol L=(L_x,L_y,L_z)$. The value of this constant of motion fix the relation of the angles $\theta$ and $\varphi$
along a trajectory. For example, for $\ell_y=0$, $\alpha_{\rm y}=\pi$, the relation (\ref{ypm}) gives the following expression for  $\theta(\varphi)$ (or $\varphi(\theta)$ )
\begin{equation}\label{tphi}
\cot{\theta}=-\frac{\ell_x}{\ell_z}\cos{\varphi}\,.
\end{equation}
However, as mentioned before, this relation also breaks when $\ell=\ell_z$, since
 in this case $\ell_x=0$ and therefore $\theta=\pi/2$.
Note that, from expression (\ref{ypm2}), it is clearly seen that $Y^\pm$ are polynomial in the momentum variables.

In this subsection, we have arrived at  an obvious result: the Hamiltonian ${\bf L}^2$  has the symmetries $L_z$ and $Y^{\pm}$ or equivalently $L_x$, $L_y$ and $L_z$. Nevertheless, the basis $L_z$, $Y^{\pm}$ will be the most adequate to write the symmetry algebra.

\end{itemize}

\subsect{Trajectories of the Perlick system type I}

In summary, once fixed the values of the constants of motion $E,\ell,\ell_z$, the new constants of motion
$X^\pm$ and $Y^\pm$
determine the relation between the coordinates $r$-$\theta$ and $\theta$-$\varphi$, respectively.
In this way, one can obtaine all the trajectories of the system. 
Now, in order to characterize each trajectory we will use the properties
of the ladder and shift functions.

The ladder functions, $B^\pm$ of (\ref{hfacb}), and
the shift functions, $A^\pm$ of (\ref{htfaca}),  can be expressed as
\begin{equation}
\begin{array}{l}
B^\pm(\xi,p_\xi) = \left(E+\frac12\left(\frac1{\ell^2}-\k\, \ell^2\right)\right)^{1/2}
e^{\pm i\, b(\xi,p_\xi)}\,,
\\[2.ex]
A^\pm(\theta,p_\theta)=\left(\ell^2-\ell_z^2\right)^{1/2}e^{\pm i\, a(\theta,p_\theta)}
\,,
\end{array}
\end{equation}
where $b(\xi,p_\xi)$ and $a(\theta,p_\theta)$ are real phase functions that
depend also on the constants of motion $E,\ell$ and $\ell_z$.
The effective Hamiltonian $H_\xi$
given in (\ref{hperlik3a}) depends of the variables $(\xi,p_\xi)$. When the energy $E$ satisfies the restrictions (\ref{energycondition}) the motion is periodic between two turning points.
The variables $(\theta,p_\theta)$ are described by the effective
Hamiltonian $H_\theta$ given in (\ref{hperlikL1}). For $\ell_z\neq 0$, the
motion of these variables will be periodic, the range of $\theta$ is
determined by its corresponding turning points. As a consequence, the functions
$b(\xi,p_\xi)$ and $a(\theta,p_\theta)$ will also be periodic. Now, taking into
account the constants of motion $X^\pm$  as given in (\ref{xpm}) the phases 
of $A^\pm$, $B^\pm$ and $X^\pm$ are related 
as follows
\begin{equation}\label{abalpha}
m\,a(\theta,p_\theta) -n\, b(\xi,p_\xi)  = \alpha_{\rm x},
\qquad
\xi_1\leq \xi\leq \xi_2,\quad \theta_1\leq \theta \leq \theta_2 \,.
\end{equation}
This equation fixes the relation of the variables $(\xi,p_\xi)$ and $(\theta,p_\theta)$ along the motion. Therefore, if we differentiate (\ref{abalpha}) with respect to time,
we get
\begin{equation}
m\, \dot a(\theta,p_\theta) - n\, \dot b(\xi,p_\xi) = 0.
\end{equation}
This implies that the frequencies $\omega_\xi$ and $\omega_\theta$ are related by
\begin{equation}\label{wtx}
m\, \omega_\theta - n\, \omega_\xi   = 0.
\end{equation}

In the same way, we can write the functions $C^\pm$ and $D^\pm$  in the form
\begin{equation}
\begin{array}{l}
C^\pm(\theta,p_\theta) = (\ell^2-\ell_z^2)^{1/2}
e^{\pm i\, c(\theta,p_\theta)}\,,
\\[2.ex]
D^\pm(\varphi,p_\varphi)=\ell_z \,e^{\pm i\, d(\varphi,p_\varphi)}\,,
\end{array}
\end{equation}
where $c(\theta,p_\theta)$ and $d(\varphi,p_\varphi)$ are real phase functions.
Due to the constants of motion $Y^\pm$ these variables are related by
\begin{equation}\label{cdalpha}
c(\theta,p_\theta) - d(\varphi,p_\varphi) = \alpha_{\rm y},
\qquad
\theta_1\leq \theta \leq \theta_2 \,,\quad -\pi\leq \varphi<\pi\,.
\end{equation}
As the motion of the $(\theta,p_\theta)$ variables and the $(\varphi,p_\varphi)$
is periodic, differentiating (\ref{cdalpha}) with respect to time, we obtain
\begin{equation}\label{wtp}
\omega_\theta -  \omega_\varphi   = 0.
\end{equation}
Hence, the frequencies of the angular variables $\theta$ and $\varphi$ are equal (except for the case $\ell=\ell_z$).

In conclusion, when the energy $E$ satisfies the restrictions 
(\ref{energycondition}), the motion is bounded, and the frequencies of the 
three variables are related by (\ref{wtx}) and (\ref{wtp}). This implies that
the bounded motion is periodic and the trajectories are closed. In conclusion,
we have checked in this case that the Bertrand's theorem is satisfied.

\begin{figure}
\centering
\includegraphics[width=0.35\textwidth]{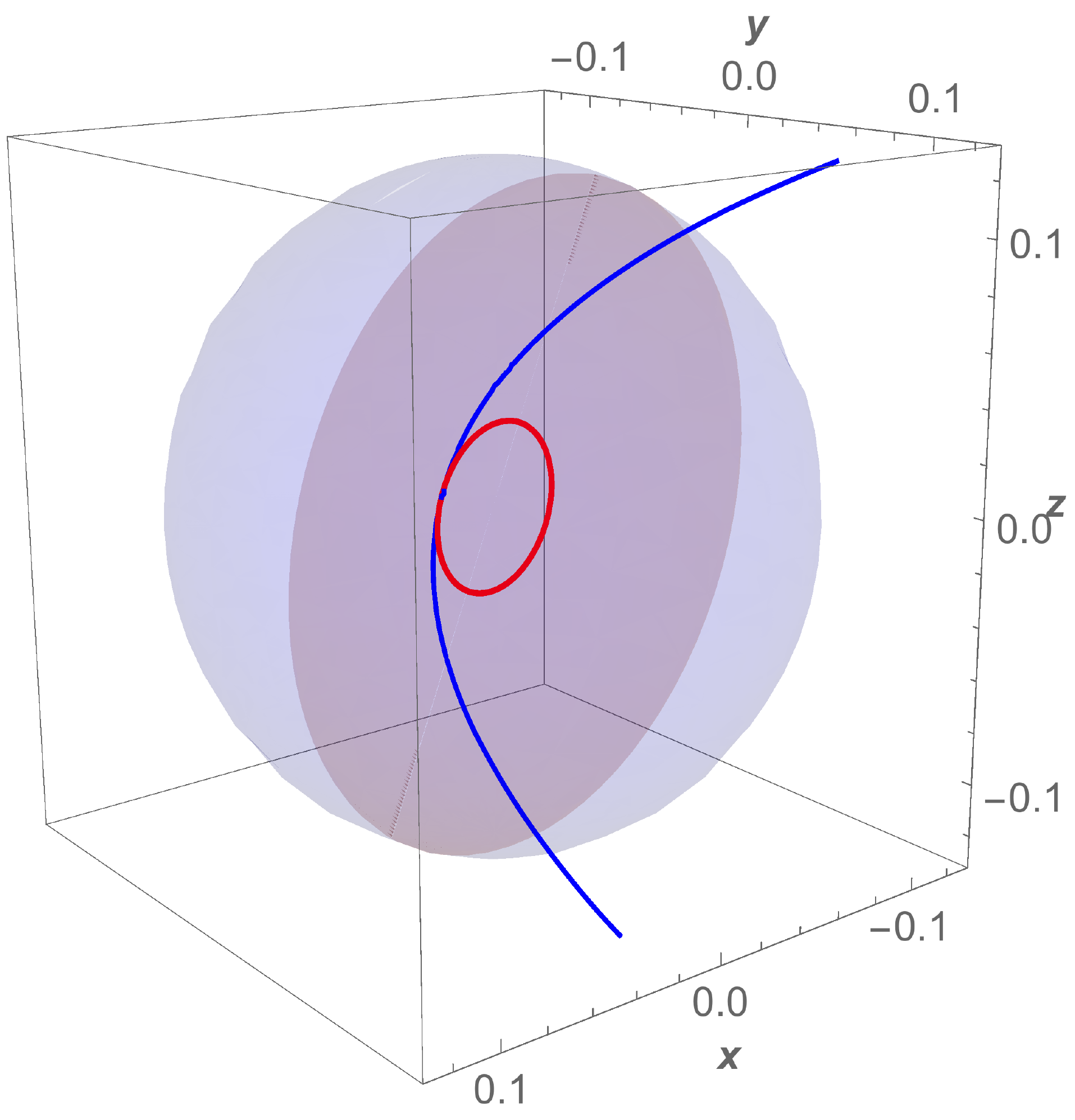}
\qquad 
\includegraphics[width=0.35\textwidth]{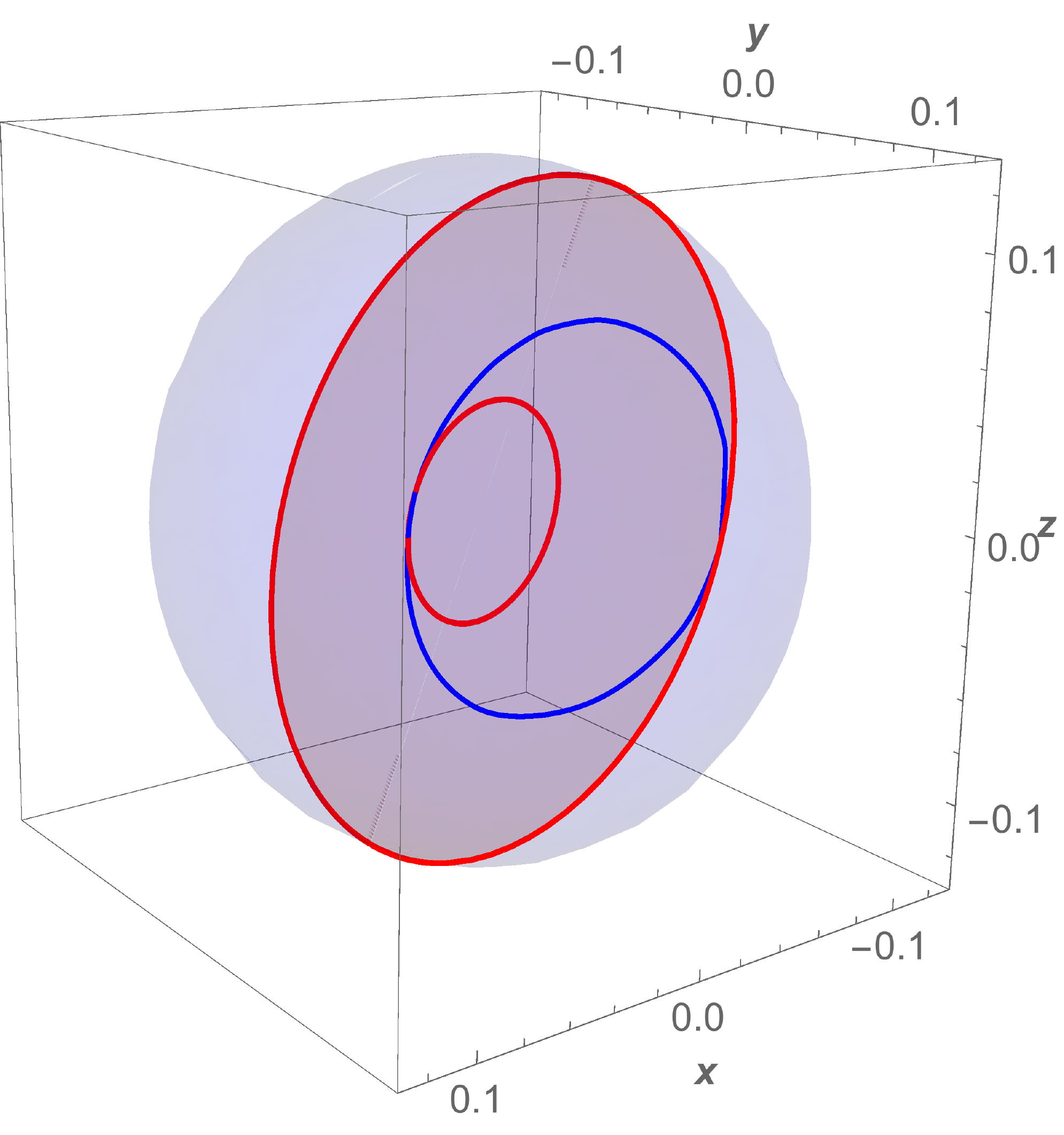}
\caption{\small Plot of the trajectories of $\beta=1$ for the values $E=-1$  (left), $E=-6$  (right), $\k=-1$, $\ell=0.25$ and $\ell_z=0.1$.
\label{trajectories1}}
\end{figure}

\begin{figure}
\centering
\includegraphics[width=0.35\textwidth]{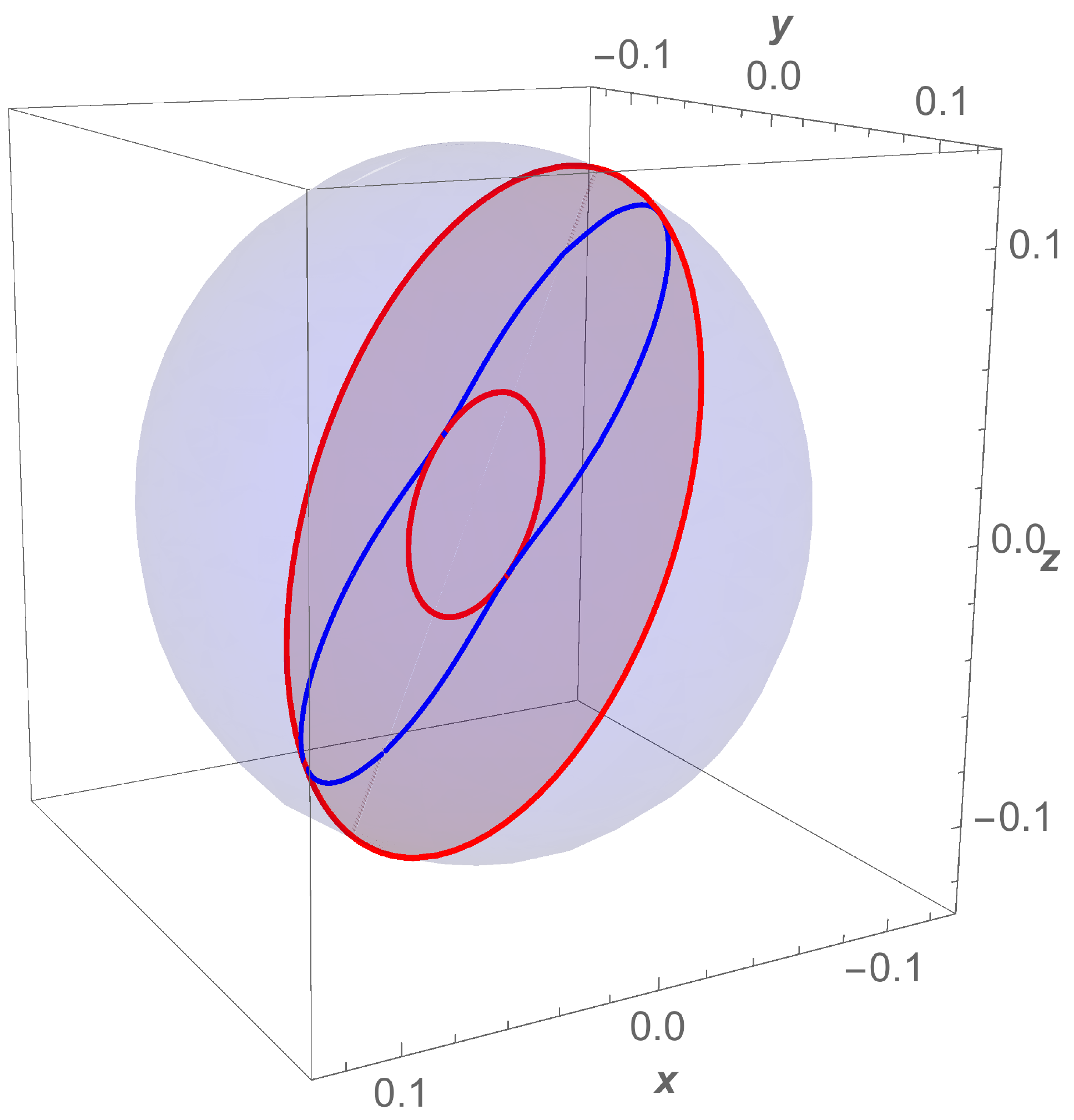}
\qquad 
\includegraphics[width=0.35\textwidth]{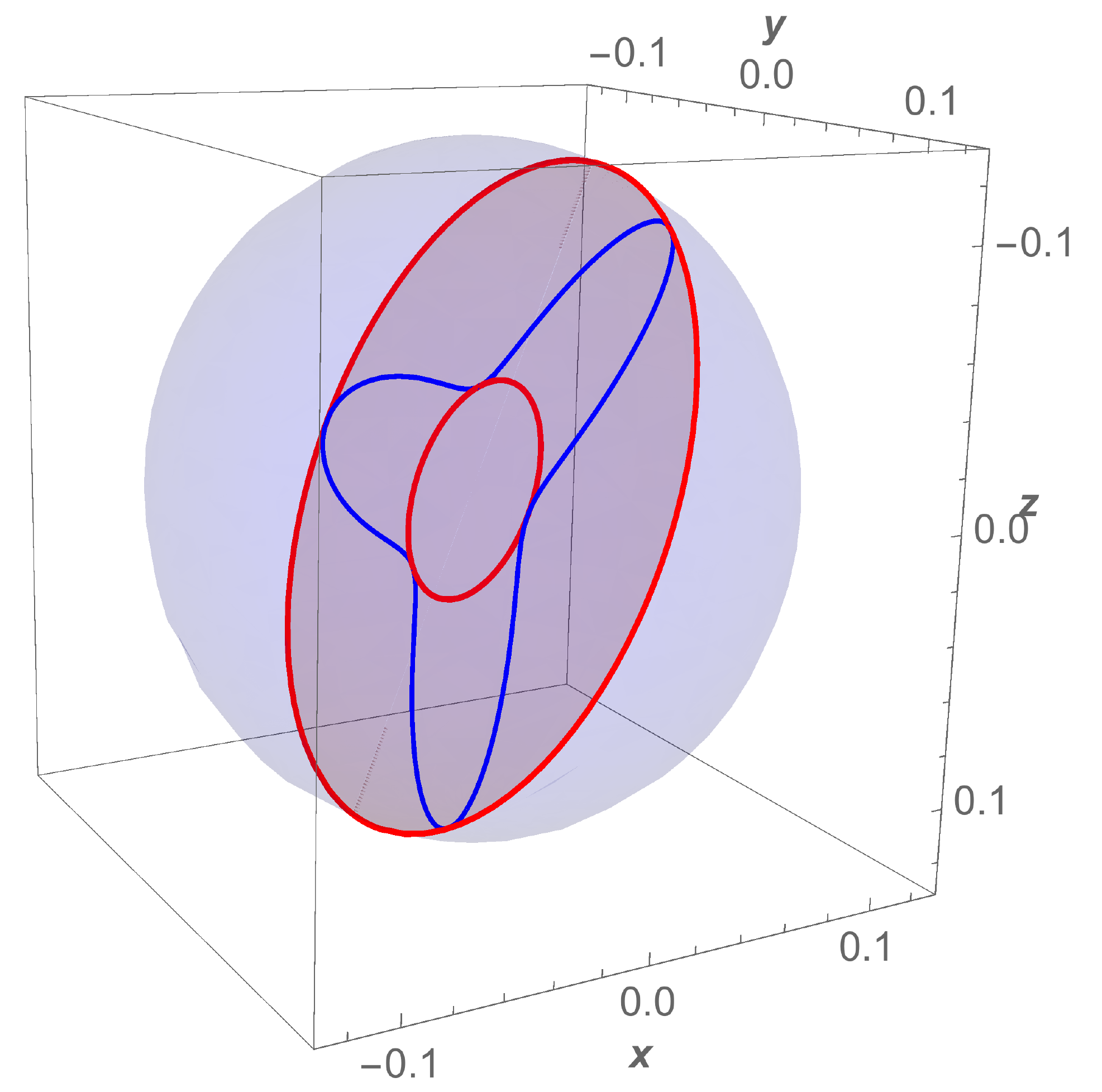}
\caption{\small Plot of the trajectories of $\beta=2$ (left),  $\beta=3$ (right) for the values $E=-6$, $\k=-1$, $\ell=0.25$ and $\ell_z=0.1$.
\label{trajectories2}}
\end{figure}

In Fig.~\ref{trajectories1} and Fig.~\ref{trajectories2} some examples of the
trajectories for different values of the energies $E$ and of  the parameter $\beta$ corresponding to bounded and unbounded motions are shown  for the case $\k=-1$. The trajectories of the initial Hamiltonian (\ref{hperlik1}) are
plotted in a three dimensional space $(r \sin \theta \cos \varphi,  r\sin \theta \sin \varphi, r \cos \theta)$, 
where $(r,\varphi,\theta)$  
are the spherical coordinates in $\mathbb R ^3$. 


\subsect{Algebra of the constants of motion for the Perlick system type I}

As we have seen in the previous sections,  we have obtained seven constants of motion
for the three dimensional Perlick system: $H, H_{\theta \varphi}, H_{\varphi}, X^{\pm}, Y^{\pm}$. Only five of these constants are independent, for example we can choose: $H, H_{\theta \varphi}, H_{\varphi}$,  $X^{+}, Y^{+}$.
Therefore, the Perlick system type I for a rational
$\beta=m/n$ is a maximally superintegrable system. It turns out that, all the aforementioned seven constants are useful to express in a simple form the algebraic structure 
defined in terms of Poisson brackets:
\begin{equation}\label{PBs}
\begin{array}{l}
\{H,\cdot\}=0\,,\qquad \{H_{\theta \varphi},Y^{\pm}\}=0\,,\qquad \{H_{\theta \varphi},X^{\pm}\}=\pm2\,i\,m\,\sqrt{H_{\theta \varphi}}\,X^{\pm},
\\[2.ex]
 \{H_{\varphi},Y^{\pm}\}=\mp\,2\,i\,\sqrt{H_{\varphi}}Y^{\pm}\,,\qquad \{H_{\varphi},X^{\pm}\}=0\,,\qquad \{Y^{+},Y^{-}\}=2\,i\,\sqrt{H_\varphi}\left(H_{\theta \varphi}-2H_{\varphi}\right),
 \\[2.ex]
 \{X^{+},X^{-}\}=i\,m\,n\,(H_{\theta \varphi}-H_{\varphi})^m\,(H_{\xi}+\dfrac12(\dfrac{1}{H_{\theta \varphi}}
 -\k\,H_{\theta \varphi}))^{n-1}(\k\,\sqrt{H_{\theta \varphi}}+\dfrac{1}{H_{\theta \varphi}^{3/2}})
 \\[2.ex]
\hskip2cm  -2\,i\,m^2(H_{\theta \varphi}-H_{\varphi})^{m-1}\,(H_{\xi}+\dfrac12(\dfrac{1}{H_{\theta \varphi}}-\k\,H_{\theta \varphi}))^{n}\,\sqrt{H_{\theta \varphi}},
 \\ [2.ex]
 \{X^{\pm},Y^{\pm}\}=\mp\dfrac{i\,m}{\sqrt{H_{\theta \varphi}}+\sqrt{H_{\varphi}}} X^{\pm}Y^{\pm},\qquad  \{X^{\pm},Y^{\mp}\}=\mp\dfrac{i\,m}{\sqrt{H_{\theta \varphi}}-\sqrt{H_{\varphi}}}X^{\pm}Y^{\mp}\,.
\end{array}
\end{equation}
In this way, we have found the algebra of the constants of motion for any two coprime
integers $m$ and $n$ and any value of the constant $\k$. The corresponding polynomial 
algebras can be found as in \cite{negro14}.

The  complex 
constants of motion $X^{\pm}$ and $Y^{\pm} $ allow to get real constants: 
$Y_R= {\rm Re}(Y^\pm), Y_I= {\rm Im}(Y^\pm)$, 
$X_R= {\rm Re}(X^\pm), X_I={\rm Im}(X^\pm)$, which will close a real algebra.  
Besides, for $\beta=1, \k=0$,  the complex constants of motion $X^\pm$  
can be expressed in terms of the 
angular momentum ${\bf L}=(L_x,L_y,L_z)$ and Runge-Lenz  
${\boldsymbol {\cal A}}=({ \cal A}_x,{ \cal A}_y, { \cal A}_z)$ vectors,
\begin{equation}\label{XLA}
X^{\pm}={\rm Re}(X^\pm)\pm
i\,{\rm Im}(X^\pm)=\dfrac{{ \cal A}_z}{\sqrt{2}}\pm i \dfrac{({\bf L} \times 
{\boldsymbol { \cal A}} )_z}{\sqrt{2}\,\ell}\,,
\end{equation}
where
\begin{equation}\label{LA}
{\boldsymbol { \cal A}}={\bf p} \times {\bf L} - \hat {\bf r}\,.
\end{equation}
Here, we have used the expressions of ${\bf r},{\bf p}, {\bf L}$ in spherical coordinates.  
Notice that in the context of the analysis of the Perlick system, a generalization of the  Runge-Lenz vector has been proposed in ref. \cite{orlando09}.
In the next section we will consider the motion in a plane corresponding to special case $\theta=\pi/2$.

\sect{Special case of the Perlick system type I}

In this section, we will study the motion of the Perlick system type I in a plane, in this
way we want to simplify the relevant constants of motion and the expression of
the trajectories determined by such constants.

Let us take $\theta=\pi/2$; with this election the motion is limited to $x$-$y$ plane and the angular momentum along the $z$-axis is equal to the total angular momentum: 
${\bf L}=(0,0,L_z)$ $(\ell= \ell_z )$. 
The corresponding Hamiltonian in the remaining $r, \varphi$ variables is
\begin{equation}\label{hperlik1s}
\widetilde H^{\pm}=\beta^2 (1+K\, r^2)\,\dfrac{p_r^2}{2}+\dfrac{{p_{\varphi}}^2}{2\,r^2}\pm\dfrac{1}{r}\,\sqrt{1+K\,r^2}\,.
\end{equation}
Now, we have two trivial constants of motion: the total energy $E$ and $\ell_z$.
Following the same procedure as in Sections 2, 3, with the change of the variables $(r,p_r)$ by $(\xi,p_\xi)$, we get the Hamiltonian
\begin{equation}\label{hperlik33ab}
H(\xi,\theta,\varphi)=\beta^2 \,\dfrac{p_\xi^2}{2}
+\dfrac{{p_{\varphi}}^2}{2}\frac{1}{\sk^2(\xi)}-\frac{1}{\tk(\xi)}\,,
\end{equation}
with the same range of the variable $\xi$ as specified in Section 2.  In this case the effective Hamiltonian $H_\xi$ is given by  
\begin{equation}\label{hperlik3ab}
H_\xi=\beta^2 \,\dfrac{p_\xi^2}{2}
+\dfrac{{\ell_z}^2}{2}\frac{1}{\sk^2(\xi)} -\frac{1}{\tk(\xi)}=\beta^2 \,\dfrac{p_\xi^2}{2}+V_{\rm eff}({\xi})\,.
\end{equation}
It is easy to find two additional constants of motion $Z^\pm$ for the Hamiltonian (\ref{hperlik33ab}): 
\begin{equation}\label{Z}
\{H,Z^{\pm}\}=0\,,\qquad Z^{\pm}=(D^{\pm})^m(B^{\mp})^n\,,
\end{equation}
where
\begin{equation}\label{hfacbs}
B^{\pm}=\frac{1}{\sqrt{2}} \,\left(\mp\,i\,{\beta }\,{p_\xi}+\frac{\ell_z}{\tk{(\xi)}}- \dfrac{1}{\ell_z}\right)\,,\qquad D^{\pm}=\sqrt{H_{\varphi}}\,e^{\mp i\,\varphi}\,,
\end{equation}
and 
\begin{equation}\label{hfacs}
H_\xi=B^+ B^-+\lambda_\xi\,,\qquad \lambda_\xi=-\dfrac12\left(\dfrac{1}{\ell_z^2}-\k\,\ell_z^2\right)\,,
\end{equation}
\begin{equation}\label{hfads}
H_\varphi=p_\varphi^2=D^+ D^-\,.
\end{equation}

These constants of motion $Z^{\pm}$ have complex values denoted by
\begin{equation}\label{zconst}
Z^{\pm}=q_z\, e^{\pm i\,\varphi_z}\,,
\end{equation}
where $-\pi\leq \varphi_z< \pi $ and $0\leq q_z<\infty$. The modulus $q_z$ has a value 
determined by the other constants of motion $E$ and $\ell_z$,
\begin{equation}\label{zq}
q_z=|Z^{\pm}|=\ell_z^m \left(E+\dfrac{1}{2}\,\left(\dfrac{1}{\ell_z^2}-\k\,\ell_z^2\right)\right)^{n/2}\,.
\end{equation}

The functions $B^\pm$ and $D^\pm$ can be written as
\begin{equation}
\begin{array}{l}
B^\pm(\xi,p_\xi) = \left(E+\dfrac{1}{2}\,\left(\dfrac{1}{\ell_z^2}-\k\,\ell_z^2\right)\right)^{1/2}
e^{\pm i\, b(\xi,p_\xi)}\,,
\\[2.5ex]
D^\pm(\varphi,p_\varphi)=\ell_z \,e^{\pm i\, d(\varphi,p_\varphi)}\,,
\end{array}
\end{equation}
where $b(\xi,p_\xi)$ and $d(\varphi,p_\varphi)$ are the corresponding real phase functions.
When the energy $E$ satisfies the restriction (\ref{energycondition}),
the bounded motion of the variables $(\xi,p_\xi)$ is periodic. Besides, the motion of the
variables $(\varphi,p_\varphi)$ is also periodic, due to the angular character of $\varphi$. 
Then, substituting in (\ref{zconst}) we get
\begin{equation}
 m\, d(\varphi,p_\varphi) - n\, b(\xi,p_\xi) = \varphi_{\rm z},
\qquad
\xi_1\leq \xi \leq \xi_2 \,,\quad -\pi\leq \varphi<\pi\,.
\end{equation}
Hence, after differentiating with respect to the time we obtain the relation
between the frequencies of the two variables:
\begin{equation}\label{frpx}
 m\, \omega_\varphi  - n\, \omega_\xi = 0 \,.
\end{equation}

\begin{figure}
\centering
\includegraphics[width=0.29\textwidth]{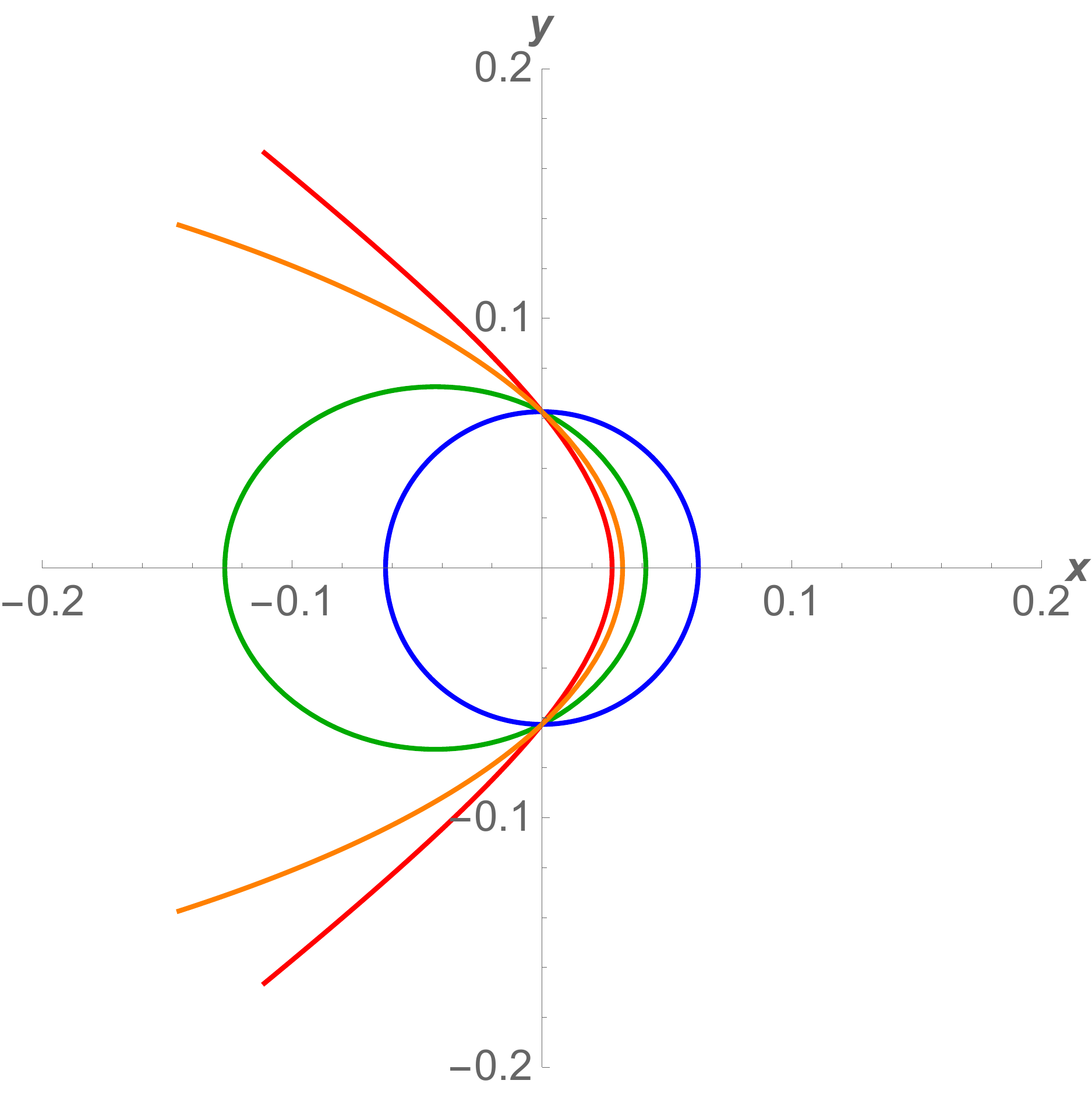}
\qquad 
\includegraphics[width=0.29\textwidth]{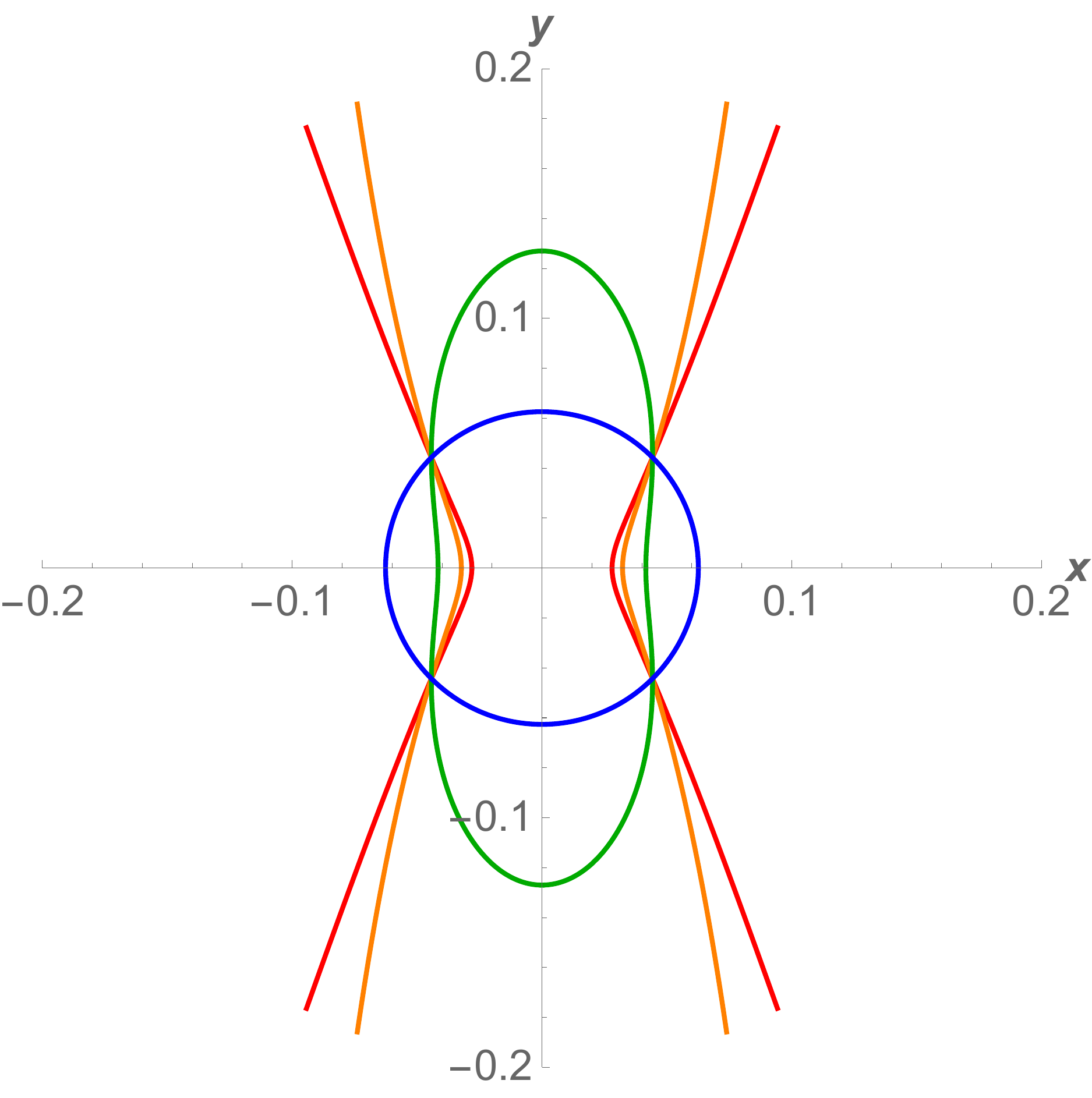}
\qquad 
\includegraphics[width=0.29\textwidth]{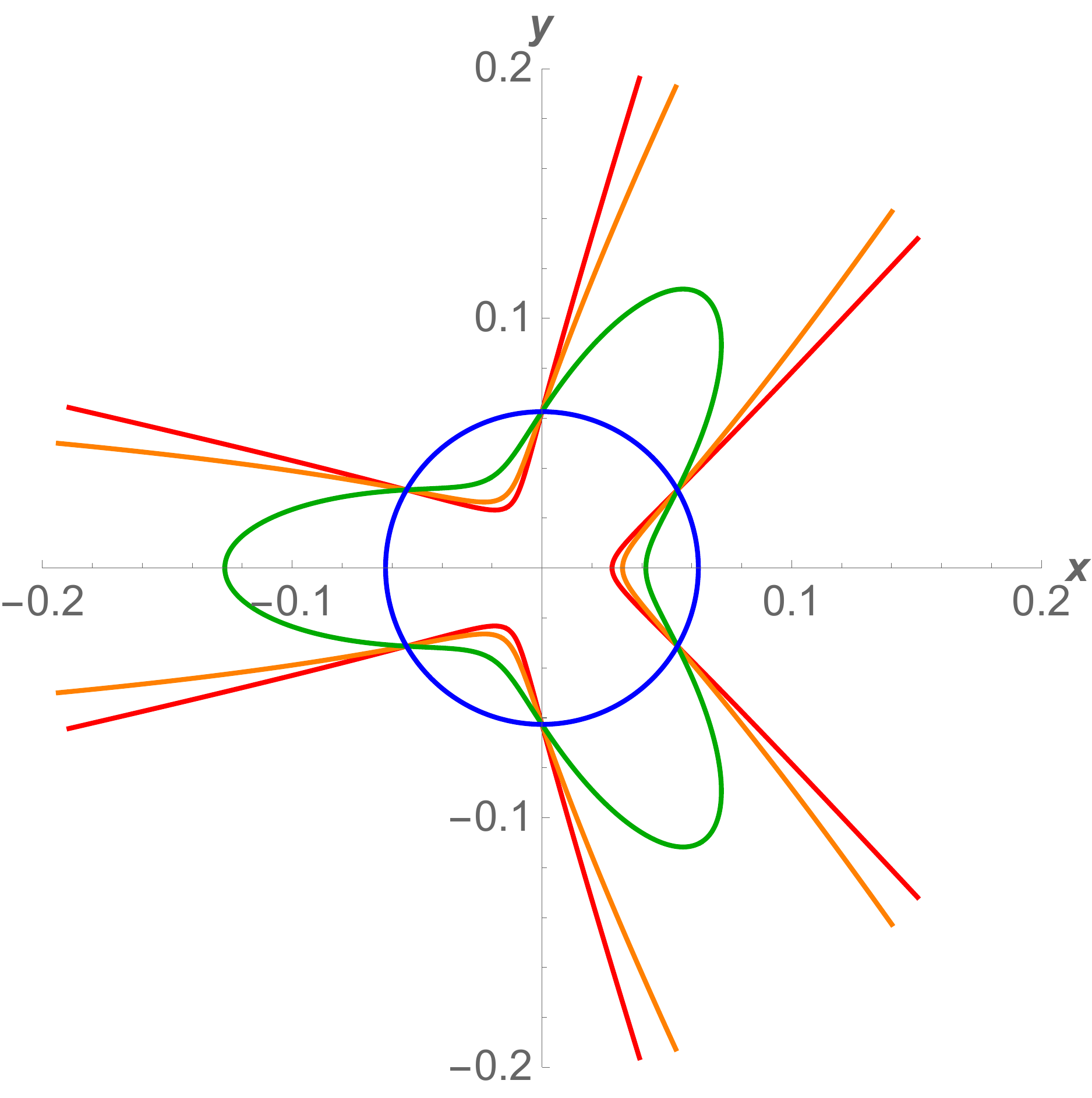}
\caption{\small Plot of the trajectories of $\beta=1$ (left), $\beta=2$ (center), $\beta=3$ (right) for the values $E=-8.03$ (minimum energy of the potential), $E=-6, E=-1, E=4$. For  $\beta=1$, respectively. Here we have chosen $\k=-1$, $\ell=0.25$ and $\ell_z=0.1$.
\label{trajectories1s}}
\end{figure}

\begin{figure}
\centering
\includegraphics[width=0.44\textwidth]{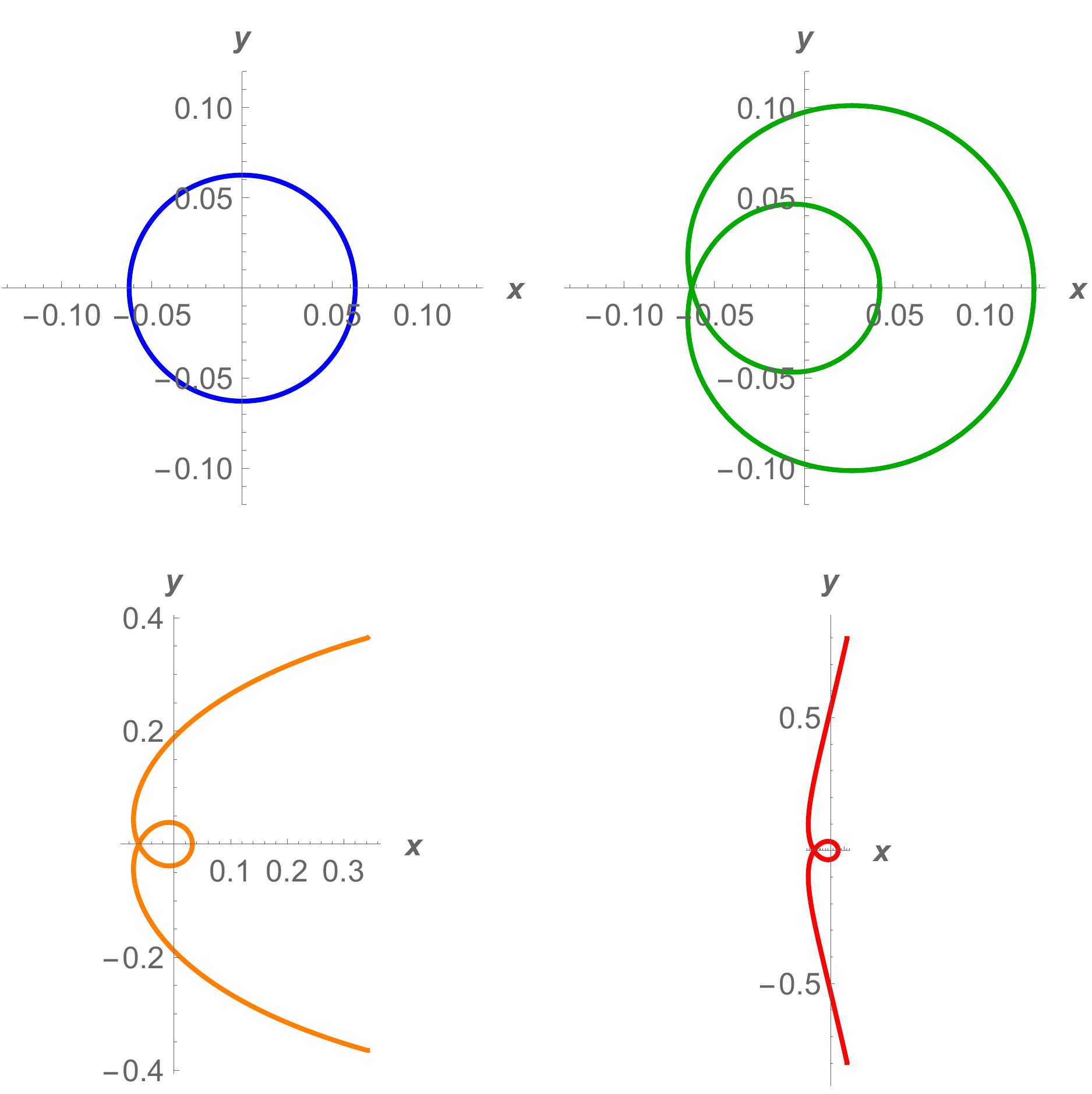}
\qquad 
\includegraphics[width=0.44\textwidth]{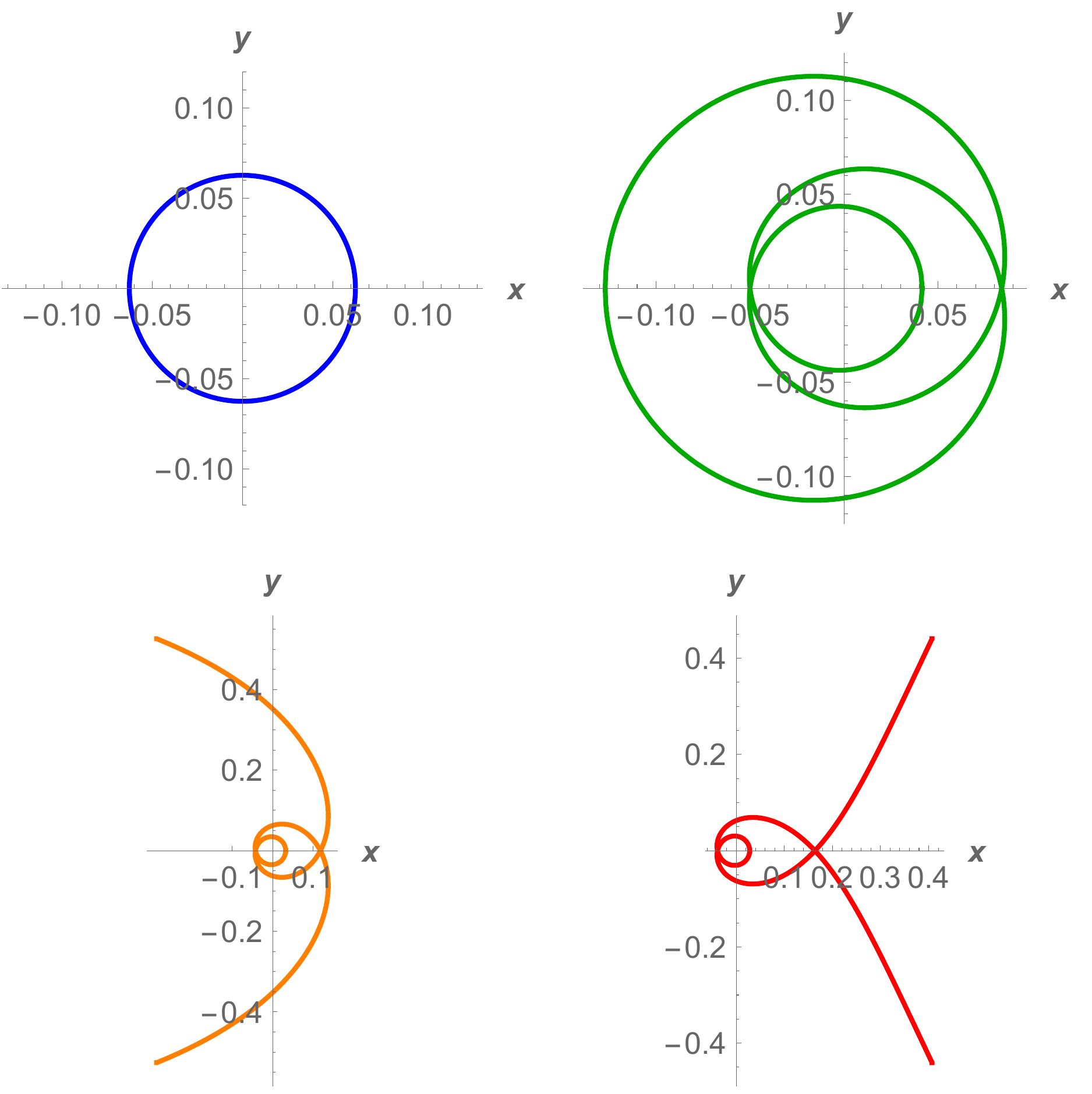}
\caption{\small Plot of the trajectories of $\beta=1/2$ (left),  $\beta=1/3$ (right) for the values $E=-8.03$ (minimum energy of the potential), $E=-6, E=-1, E=4$, corresponding to circular, elliptic, parabolic and hyperbolic orbits, respectively. Here we have chosen $\k=-1$, $\ell=0.25$ and $\ell_z=0.1$.
\label{trajectories2s}}
\end{figure}

\begin{figure}
\centering
\includegraphics[width=0.29\textwidth]{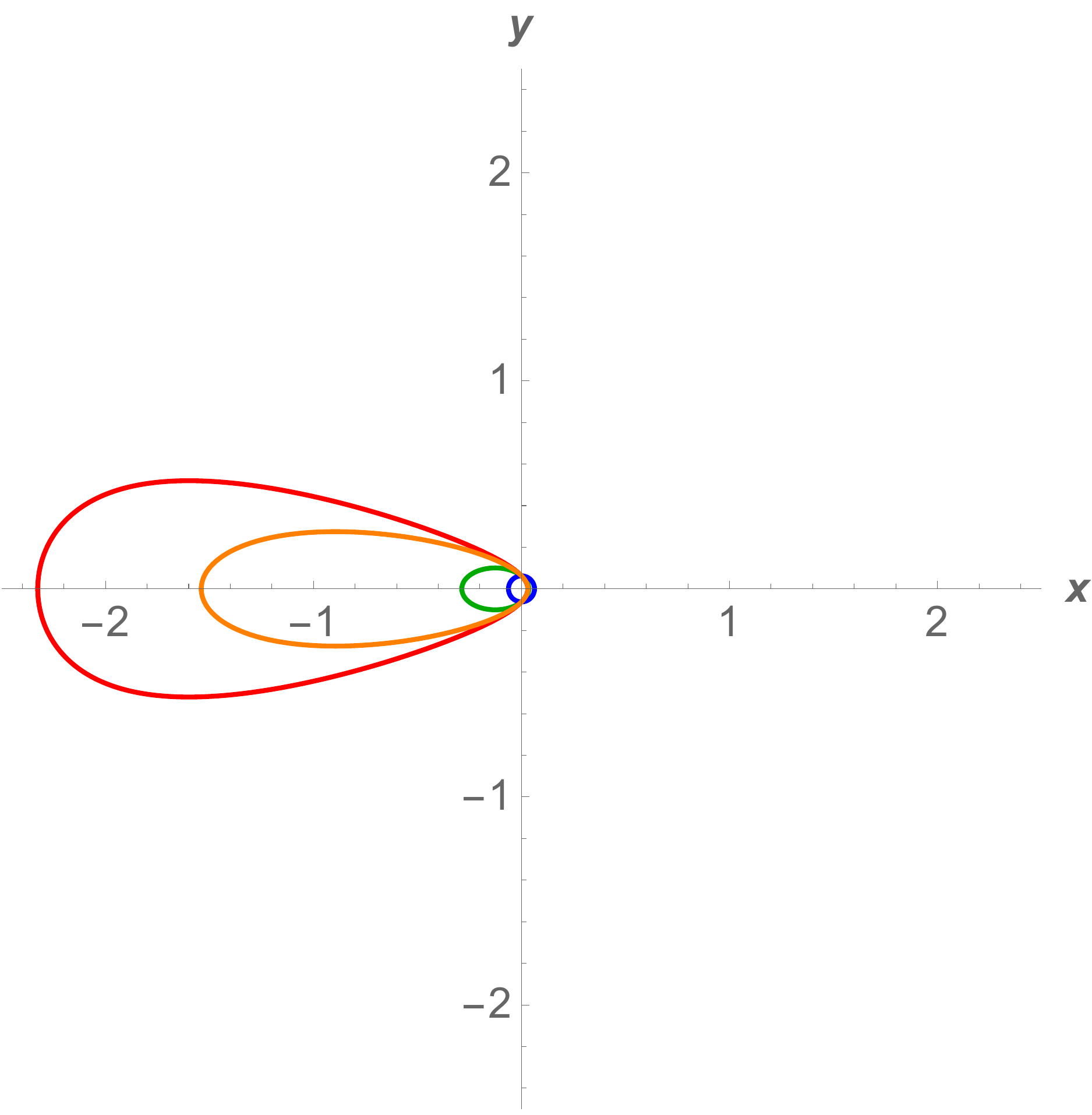}
\qquad 
\includegraphics[width=0.29\textwidth]{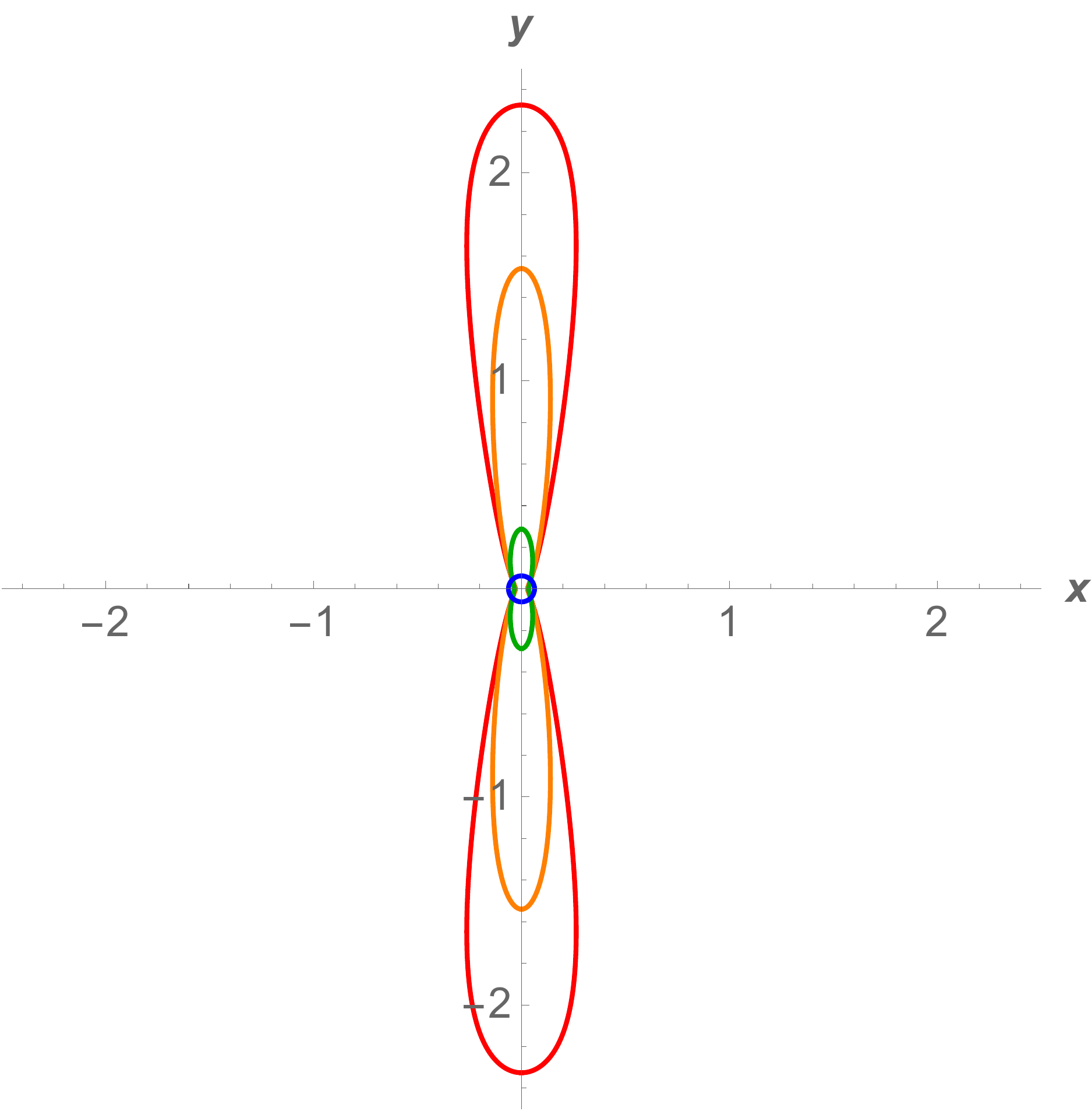}
\qquad 
\includegraphics[width=0.29\textwidth]{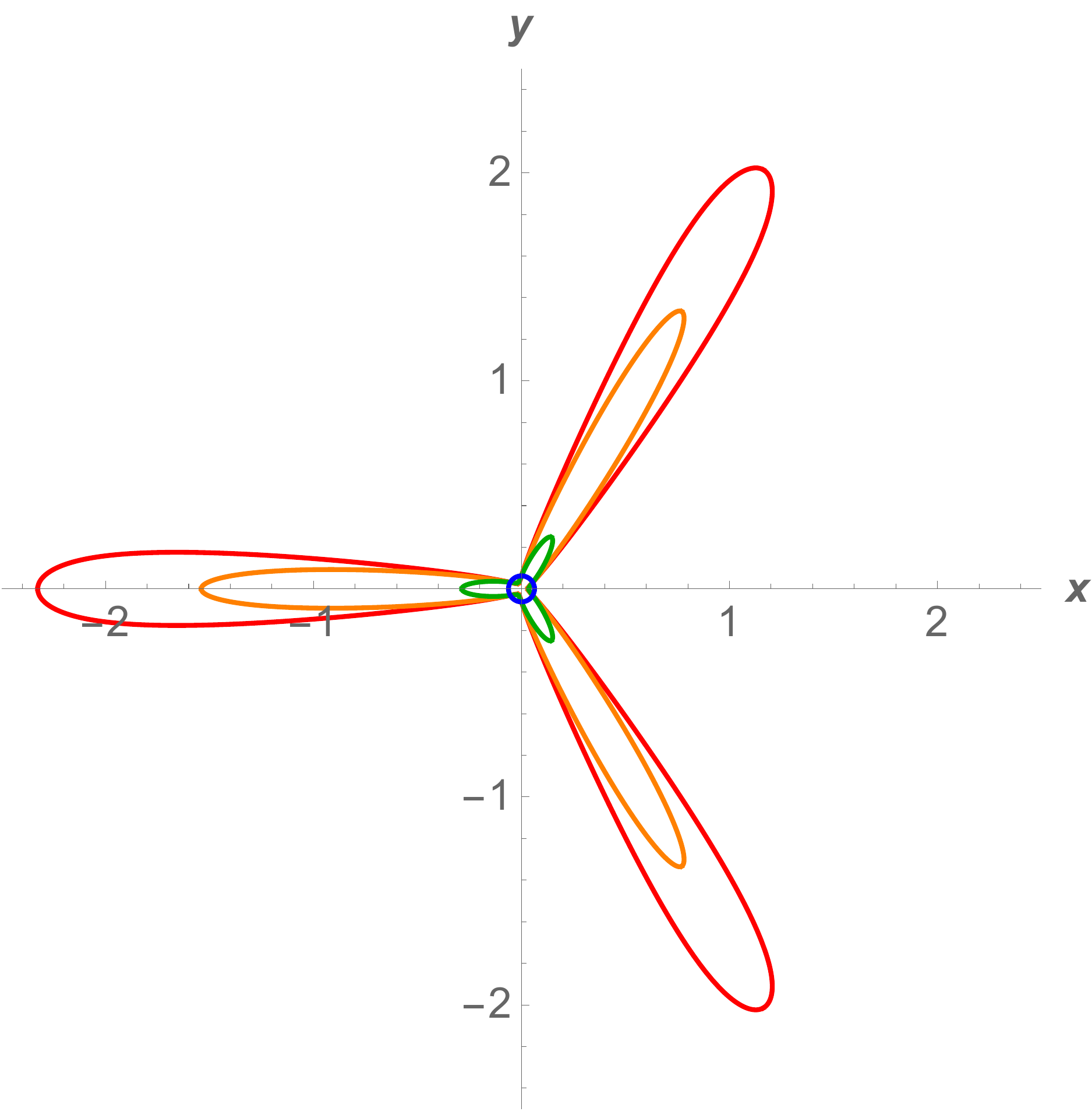}
\caption{\small Plot of the trajectories of $\beta=1$ (left), $\beta=2$ (center), $\beta=3$ (right) for the values $E=-7.96875$ (minimum energy of the potential), $E=-3, E=0, E=1$. Here we have chosen $\k=1$, $\ell_z=0.25$.
\label{trajectories1ss}}
\end{figure}

\begin{figure}
\centering
\includegraphics[width=0.44\textwidth]{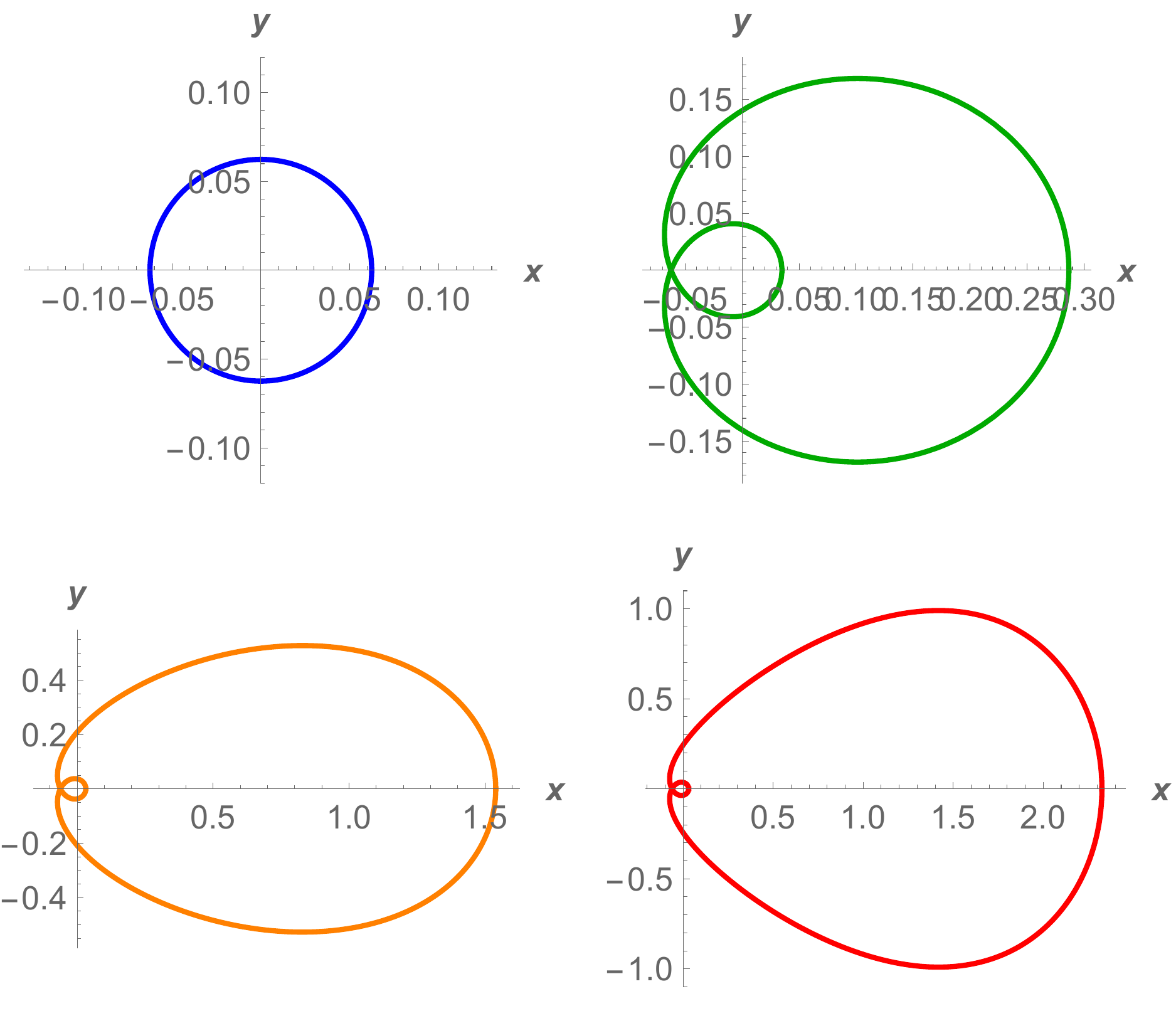}
\qquad 
\includegraphics[width=0.40\textwidth]{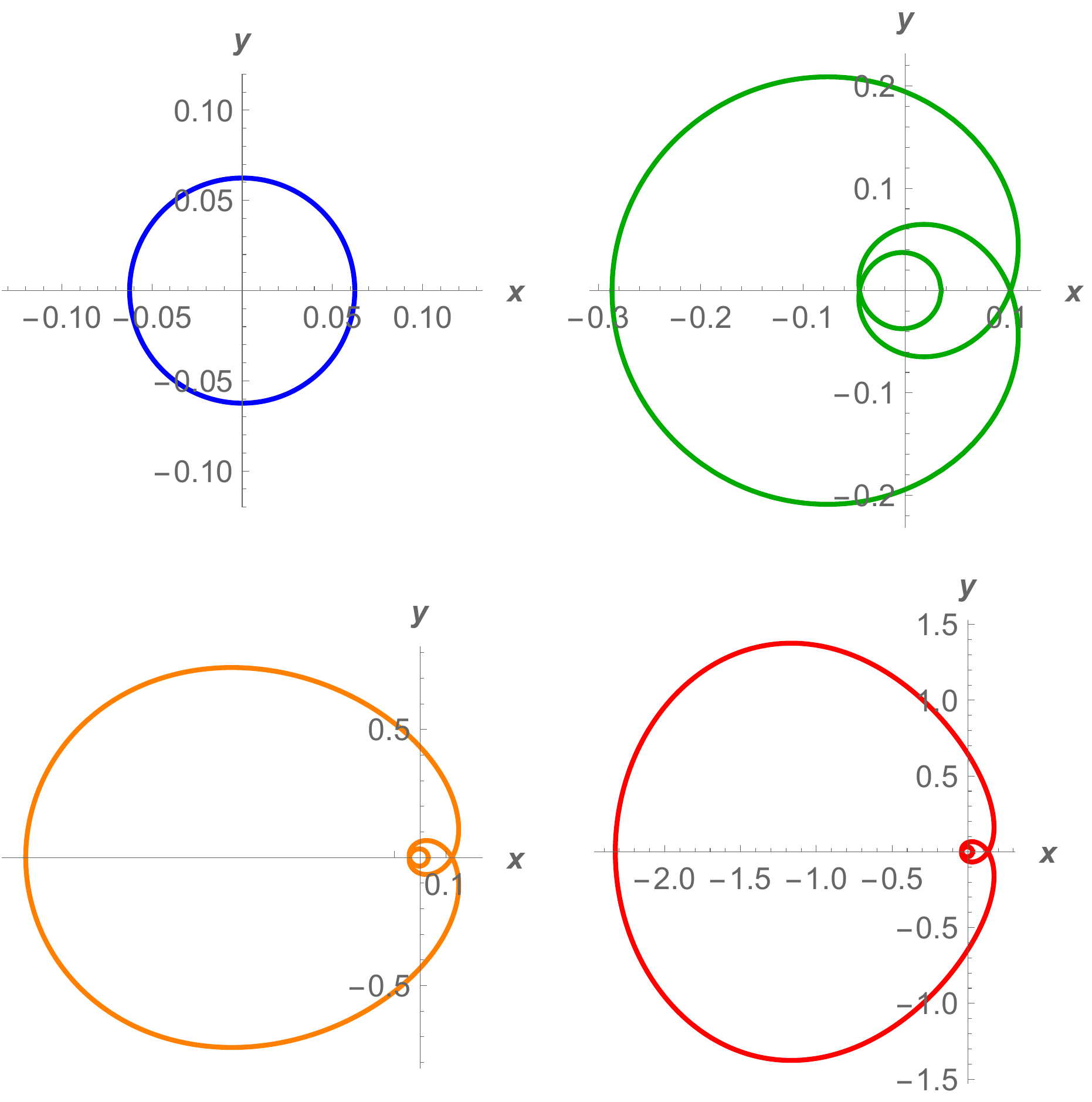}
\caption{\small Plot of the trajectories of $\beta=1/2$ (left),  $\beta=1/3$ (right) for the values $E=-7.96875$ (minimum energy of the potential), $E=-3, E=0, E=1$. Here we have chosen $\k=1$, $\ell_z=0.25$.
\label{trajectories2ss}}
\end{figure}

Let us write the equation of the constant of motion $Z^{\pm}$,
\begin{equation}\label{zqeq}
(\ell_z\,e^{\mp i\,\varphi})^m\,
(\mp i \,\dfrac{\beta}{ \sqrt{2}}\,{p_\xi}+ \dfrac{\ell_z}{\sqrt{2}}\dfrac{1}{\tk(\xi)}- \dfrac{1}{\ell_z\,\sqrt{2}})^n
=q_z\, e^{\pm i\,\varphi_z}\,.
\end{equation}
Taking the real and imaginary parts of this equality, we get  two equations
\begin{equation}\label{zqeqre}
 \dfrac{\ell_z}{\sqrt{2}}\frac{1}{\tk(\xi)}- \dfrac{1}{\ell_z\,\sqrt{2}}
= \left(\dfrac{q_z}{\ell_z^m}\right)^{1/n }\cos{\left(\frac{1}{n}\varphi_z+\frac{m}{n}\varphi\right)}\,,
\end{equation}
\begin{equation}\label{zqeqim}
- \dfrac{\beta}{\sqrt{2}}\,{p_\xi}= \left(\dfrac{q_z}{\ell_z^m}\right)^{1/n }\sin{\left(\frac{1}{n}\varphi_z+\frac{m}{n}\varphi\right)}\,.
\end{equation}
From
(\ref{zqeqre}), we obtain the relation between $\xi$ and $\varphi$ along the trajectories \cite{orlando09}:
\begin{equation}\label{zqeqre2}
\cos{\left(\frac{1}{n}\varphi_z+\dfrac{m}{n}\varphi\right)}=\frac{\dfrac{\ell_z^2}{\tk(\xi)}-1}{\sqrt{2\,E\,\ell_z^2+1-\k\,\ell_z^4}}\,.
\end{equation}
Relation (\ref{zqeqre2}) becomes a well known conic section equation for the values $\k=0$ , 
$\beta=m/n=1$ and $\varphi_z=0$:
\begin{equation}\label{conicsec}
\frac{\alpha}{\xi}=1+\varepsilon\,\cos{\varphi}\,,
\end{equation}
where $\varepsilon^2=2\,E\,\ell_z^2+1$ (eccentricity)  and $\alpha=\ell_z^2$ (semi-latus rectum)
 \cite{goldstein}.  In this case, the problem is reduced to the Kepler-Coulomb system in the Euclidean plane. If $\k=-1$ ($\k=1$) and $\beta=1$, then we get the conic section equation in the hyperboloid (sphere)  \cite{ranada05}: 
 \begin{equation}\label{conicsechsp1}
\k=-1\,,\quad {(\rm Hyperbolic)}\qquad \dfrac{\alpha}{\tanh{\xi}}=1+\sqrt{\varepsilon^2+\alpha^2}\,\cos{\varphi}\,,
\end{equation}
\begin{equation}\label{conicsechsp2}
\k=1\,,\quad {(\rm Spherical)}\qquad  \dfrac{\alpha}{\tan{\xi}}=1+\sqrt{\varepsilon^2-\alpha^2}\,\cos{\varphi}\,.
\end{equation}
So, we may say that 
the equation (\ref{zqeqre2})  can be considered as a generalized conic section equation.
Taking into account the relation (\ref{conicsechsp1})  for $\k=-1$ and the relation (\ref{conicsechsp2})  for $\k=1$
different examples of trajectories  $(\xi\,\cos \varphi,  \xi\,\sin \varphi)$, 
where $(\xi,\varphi)$ are polar coordinates, are plotted in Figs.~\ref{trajectories1s}-\ref{trajectories2ss}. In these figures there are represented bounded and unbounded trajectories, according to the values of the energy $E$. In the case of bounded motion, due to the above relation of the frequencies
(\ref{frpx}) the motion must be periodic and the trajectories  are closed.

In the special case  where $\k=0$ and $\beta=m/n=1$, the constants of motion $Z^{\pm}$ can also be expressed in terms of the  Runge-Lenz vector ${\boldsymbol {\cal A}}=({ \cal A}_x,{\cal A}_y,0)$,
\begin{equation}\label{ZLA}
Z^{\pm}={\rm Re}(Z^\pm)\pm
i\,{\rm Im}(Z^\pm)=\dfrac{ { \cal A} _x}{\sqrt{2}} \mp i \dfrac{{ \cal A} _y}{\sqrt{2}}\,.
\end{equation}
For the other cases ($\k=\pm 1$), the constants of motion $Z^{\pm}$  can be written in terms of a type of generalized Runge-Lenz vector \cite{orlando09}.
Note that, the constants $H, H_\varphi, Z^\pm$ close a similar algebra as in the general case.

\sect{Conclusions and remarks}

In this paper, we have started an algebraic approach to the study of  Perlick's systems, the family of maximally superintegrable systems discovered in 1992, that represent the most general extension of the classical Bertrand systems to curved (though conformally flat) N-dimensional (Riemannian) manifolds. 
We would summarize our main new findings as follows.
On one hand, we have given a thorough description  of the algebraic and geometrical properties of the classical Perlick system  of type I by means of the factorization approach: we have identified both the shift and the ladder functions for the radial as well as for the angular Hamiltonian. Henceforth, we have been able to  identify the full set of  constants of motion and their related Poisson algebra, and unveiled the intimate connection with the geometric features of the orbits provided by superintegrability. 
On the other hand, we have been able to single out the role played by the two parameters that characterize the system:  the real parameter $K$, entering both in the metrics and in the potential,   is related with the compact or non-compact  nature of the manifold where the motion takes place;  the rational parameter $\beta$,  which does not appear in the potential,  is in turn  responsible for what we would call ``the complexity" of the trajectories, namely their winding number. These features emerge in a perspicuous manner from the graphic representations of the (effective) potential as well as of the orbits, reported in a number of figures for different values of $K$ and $\beta$. 

The further steps to be accomplished are clear. (i) We have to see how our construction goes over to the quantum setting, and (ii) we have to perform a similar construction for the full Perlick family II. Work is actually in progress in both directions, and promising preliminary results have been already obtained by the authors and their collaborators. 

 \section*{Acknowledgments}

This work was partially supported by the Spanish Ministerio de Econom\1a y Competitividad    (MINECO) under project  MTM2014-57129-C2-1-P and Junta de Castilla y Le\'on (VA057U16).  \c{S}.~Kuru acknowledges Ankara University and the
warm hospitality at the Universidad de Valladolid, Dept.\
de F\'isica Te\'orica,  where this work has been done. O. Ragnisco wishes to thank all the
colleagues at Dept. de F\'{\i}sica Te\'orica, 
for their warm hospitality during his stay.



\vfill

\end{document}